\documentclass[11pt]{article}%
\usepackage{bbm}
\usepackage[latin9]{inputenc}
\usepackage{setspace}
\usepackage{amsmath}
\usepackage{amssymb}
\usepackage{amsfonts}
\usepackage{graphicx}
\usepackage{natbib}
\usepackage{xurl}
\usepackage[margin=1.15in]{geometry}
\usepackage[colorlinks]{hyperref}
\usepackage[svgnames]{xcolor}%
\providecommand{\U}[1]{\protect\rule{.1in}{.1in}}
\hypersetup{
pdftitle={Consensus and disagreement},
pdfauthor={Abhijit Banerjee and Olivier Compte},
citecolor=DarkBlue,
bookmarksnumbered=true,
urlcolor=Indigo,linkcolor=DarkBlue
}

\definecolor{dark-red}{rgb}{0.4,0.15,0.15}
\definecolor{dark-blue}{rgb}{0.15,0.15,0.75}
\definecolor{medium-blue}{rgb}{0,0,0.5}
\begin{document}

\begin{titlepage}
\title{Consensus and Disagreement: \\{\Large Information Aggregation under (not so) Naive Learning\thanks{This
paper was previously entitled \textquotedblleft Information Aggregation under
(not so) Naive Learning".}}}
\author{Abhijit Banerjee and Olivier Compte\thanks{Banerjee: \textit{MIT}. Compte:
\textit{Paris School of Economics}, 48 Boulevard Jourdan, 75014 Paris and \textit{Ecole des Ponts Paris Tech},
\href{mailto:olivier.compte@gmail.com}{\color{dark-blue}olivier.compte@gmail.com}. }}
\date{November 2023}
\maketitle
\begin{abstract}
We explore a model of non-Bayesian information aggregation in networks. Agents
non-cooperatively choose among Friedkin-Johnsen type aggregation rules to
maximize payoffs. The DeGroot rule is chosen in equilibrium if and only if
there is noiseless information transmission, leading to consensus. With noisy
transmission, while some disagreement is inevitable, the optimal choice of
rule amplifies the disagreement: even with little noise, individuals place
substantial weight on their own initial opinion in every period, exacerbating
the disagreement. We use this framework to think about equilibrium versus
socially efficient choice of rules and its connection to polarization of
opinions across groups.
\end{abstract}
\thispagestyle{empty}
\end{titlepage}

\setstretch{1.1}

\clearpage
\setcounter{tocdepth}{2}  
\tableofcontents
 \thispagestyle{empty}  \clearpage
\setcounter{page}{1}

\section{Introduction}

As of May 2020, 41\% of US Republicans were not planning to get vaccinated
against Covid-19, as compared to 4\% of
Democrats.\footnote{https://www.pbs.org/newshour/health/as-more-americans-get-vaccinated-41-of-republicans-still-refuse-covid-19-shots}
We saw similar divergences in mask-wearing, social distancing etc, which
protect against the disease. Since Covid-19 is a life-threatening ailment that
had already taken more than 3.5 million lives so far world-wide, it is hard to
think of these as being just empty gestures or entirely reflective of
different preferences, though there is surely some of that. There seems to be
rather, a different reading of the facts on the ground; for example, in a Pew
Research Center
poll,\footnote{https://www.pewresearch.org/fact-tank/2020/07/22/republicans-remain-far-less-likely-than-democrats-to-view-covid-19-as-a-major-threat-to-public-health.}
Republicans were much more likely to say that Covid-19 is not a major threat
to the health of the US population (53\% compared to 15\% of Democrats). This
goes with a general deepening in the political divide between Democrats and
Republicans in recent years.\footnote{Pew Center (2014) documents such a shift
of political values for the period 1994-2014. See also \cite{gentzkow16} and
\cite{bertrand22}}

The source of this shift is a subject of much discussion: one potential source
of change is the massive growth in the use of the internet. However the
evidence from the careful work by \cite{gentzkow11} suggests that online news
consumption is not more segregated by political leanings than other sources of
information that already existed, contrary to the concerns expressed for
example by \cite{sunstein01}.\footnote{Though \cite{guess21} suggests that the
segregation in news consumption has been increasing in recent years.} The most
segregated sources of information, according to \cite{gentzkow11} seem to be
social networks (voluntary associations, work, neighborhoods, family,
\textquotedblleft people you trust", etc), which were of course always there.
However there is evidence that online networks such as Facebook are
substantially more segregated than other social networks and as a result, news
that comes from being shared on Facebook tends to be more segregated than news
from other media sources (\cite{bakshy15}).\footnote{The Facebook news feed
turns out to be even more seggregated (\cite{levy21}).} It is true that social
media are still a relatively small, though growing, part of news consumption,
but the volume of "information" that can be quickly shared on Facebook may be
much larger than other more traditional sources. Moreover while information
was always shared through social connections, the evidence of growing
affective polarization along political lines, especially in the US
(\cite{boxell22}), raises the concern that the actual exchange of sensitive
information in the social network is increasingly confined to those with
similar views.

Given this evidence, we feel that it is worth exploring theoretically when and
why social learning on networks can lead to large and persistent
disagreements. As a starting point, we note that models of Bayesian social
learning such as \cite{acemoglu11} propose relatively weak conditions on
signals and network structure under which information is perfectly aggregated
as the network grows to be very large. More recent work, in which agents
repeatedly communicate (unlike in \cite{acemoglu11} where they communicate
only once) includes \cite{mossel15} who derive necessary conditions on the
network structure under which Bayesian learning yields consensus and perfect
information aggregation.\footnote{They build on \cite{rosenberg09} and the
literature on \textquotedblleft Agreeing to Disagree\textquotedblright that
goes back to \cite{aumann76}.} The general sense from this literature is that
convergence to a consensus is likely even when the network exhibits a
substantial degree of homophily (Republicans mostly talk to other Republicans)
as long as everyone is ultimately connected.

This \textit{Bayesian route} however requires that agents make correct
inferences based on an understanding of all the possible ways information can
transit through the network, which, at least for large networks, strains
credibility.\footnote{A Bayesian needs to think through all possible sequences
of signals that could be received as a function of the underlying state and
all the possible pathways through which each observed sequence of signals
could have reached them. As discussed in \citet[p.~1681]{alatas16}, there is
obviously an extremely large number of such pathways.
}

The alternative way to model learning on networks is to take a
\textit{non-Bayesian route}, which avoids these very demanding assumptions
about information processing by postulating a simple rule that individuals use
to aggregate own and neighbors' opinions. In recent years the economics
literature has tended to favor the DeGroot (DG) rule, where agents update
their current opinion by linearly averaging it with their neighbors' most
recent opinions. As observed by \cite{demarzo03}, who brought it into the
economics literature, the rule builds in a strong tendency towards consensus
in any connected network, even when there is high degree of homophily and
people put high weight on people like them, though convergence between those
far from each other in the network can be very slow.\footnote{Moreover as
shown by \cite{golub10}, DG has the striking property that, under some
restrictions on network structure and weights on neighbors, learning converges
to perfect information aggregation in large networks.} Faced with this force
towards consensus, \cite{friedkin90} came up with a learning rule which is
similar to DG, but allows each individual to keep putting some weight on their
own initial opinion.\footnote{\citet[p~3]{friedkin99} write, referring to the
work of \cite{degroot74} and other precursors: \textquotedblleft These initial
formulations described the formation of group consensus, but did not provide
an adequate account of settled patterns of disagreement\textquotedblright.}
This rule, for obvious reasons, does not lead to a consensus.

The first question we set out to answer here is which type of rule, i.e.,
Friedkin-Johnsen (FJ) or DG would be favored by individuals given a choice. In
other words are there good reasons to prefer rules where individuals anchor
themselves to their initial beliefs even while updating their opinions based
on what they are hearing from others?

To study this question, we start from a broad class rules in the spirit of
Friedkin-Johnsen (FJ), which includes DG and can formally be written as
\begin{equation}
y_{i}^{t}=(1-\gamma_{i})y_{i}^{t-1}+\gamma_{i}(m_{i}x_{i}+(1-m_{i})z_{i}%
^{t-1}) \tag{FJ}\label{FJ}%
\end{equation}
where $y_{i}^{t}$ is ${i}$'s belief in period ${t}$, $x_{i}$ is the initial
signal that $i$ received, correlated with some underlying state of the world
(we shall refer to $x_{i}$ as $i$'s initial opinion or \textit{seed}), and
\begin{equation}
z_{i}^{t}=\sum_{j\in N_{i}}A_{ij}y_{j}^{t}+\varepsilon_{i}^{t} \label{eqz}%
\end{equation}
is the weighted average of reports received by $i$ from his neighbors (denoted
$N_{i}$),\footnote{The matrix $A=(A_{ij})_{ij}$ defines the weight $A_{ij}$
that $i$ puts on $j$'s opinion, with $A_{ij}>0$ if and only if $j\in N_{i}$,
and $\sum_{j}A_{ij}=1$.} plus any processing or transmission error. This error
term is an important ingredient of our analysis. We assume that $\varepsilon
_{i}^{t}$ has two components, a persistent one, drawn at the start of the
process, and an idiosyncratic one, drawn at each date, though, to simplify the
exposition, much of the paper focuses on persistent errors.
When the weight $m_{i}$ is $0$, individual $i$ is using a DG
rule.\footnote{Throughout our analysis, we assume that all $\gamma_{i}$ are
strictly positive.}

Within this limited class of \textquotedblleft natural" rules, parameterized
by $\gamma_{i}$ and $m_{i}$,\footnote{We assume that the weights $A_{ij}$ are
fixed, not subject to optimization.} we allow agents full discretion in the
choice of rules and assume that each individual non-cooperatively selects
$m_{i}$ and $\gamma_{i}$ to ensure that the long-run opinion $y_{i}$ is on
average closest to the underlying state. This is in the spirit of the approach
advocated in \cite{compte18} to model mildly sophisticated
agents.\footnote{The limitation to a specific class of rules is key. Otherwise
the individually optimal way to process signals among all possible
signal-processing rules would be the Bayesian rule.}

Our results highlight the major role of errors in shaping equilibrium choices
and outcomes. Result~1 says that \textit{absent errors}, each individual
decision-maker will choose DG ($m_{i}=0$) in the Nash equilibrium of the
rule-choice game, hence there will be consensus. Moreover, we show that each
individual
will choose $\gamma_{i}$ in such a way that information is efficiently
aggregated. This result thus complements \cite{golub10} who show that when
everyone does DG (but do not choose their $\gamma_{i}$), information
aggregation in large networks is almost perfect under certain weak conditions,
but generally imperfect in finite networks.

In contrast, Result~2 shows that in the presence of any error in transmission,
each decision-maker must choose $m_{i}>0$ in equilibrium, so there will be no
consensus even in the long run. The reason is that when all the $m_{i}$ are
small (a fortiori when everyone uses DG) the errors tend to cumulate, with the
result that long-run opinions explode. Intuitively, a positive error by $i$
pushes up $i$'s opinion, which raises the opinions of others', fueling a
further rise in $i$'s opinion, and so on --We call these \textit{echo
effects}. Raising $m_{i}$ allows individuals to limit this cumulation of
errors, at the cost of potentially putting too much weight on their own seeds.
Moreover there is no way to use $\gamma_{i}$ to mitigate this problem: in fact
as long as there is no \emph{idiosyncratic} error and $m_{i}>0$ for a least
one player, $\gamma_{i}$'s play no role: long-run opinions are \textit{fully}
determined by the $m_{i}$'s. Later in the paper we show that $\gamma_{i}$ does
play an important role in controlling the effects of idiosyncratic errors, but
that does not change the need to set $m_{i}>0$. \smallskip

It should be clear that in any Nash Equilibrium of the rule choice game, there
are two sources of divergence of opinions--the errors themselves, but also the
additional divergence that comes from always putting non-zero weight on one's
initial signal (which is a choice, but one resulting from the presence of
errors). The next question is which is the main source of divergence.

Result~3 shows that at least when the variance $\varpi$ of persistent error is
close enough to zero, the second, non-mechanical, source dominates:
specifically we show that in equilibrium, the weights $m$ are comparable to
$\varpi^{1/3}$. A rough intuition goes as follows: from the perspective of
player $i$, when other players use $m_{j}\simeq m$, the cumulated error he
faces has a long-run variance of the order of $\varpi/m^{2}$. $i$ will want to
set $m_{i}$ to counterbalance this, which means at the order of $\varpi/m^{2}%
$. Therefore in equilibrium, $m\simeq O(\varpi/m^{2})$.

We then compare the extent of disagreement in any equilibrium to the social
optimum. Result~4 shows that there is too little-- equilibrium values of
$m_{i}$ are always lower than the socially optimal values. One reason is that
in setting $m_{i}$ optimally, player $i$ does not take into account the fact
that lowering $m_{i}$ raises the cumulated error faced by $j$. But
%
this is not the only reason.
In choosing $m_{i}$, $i$ trades off the fact that a higher value of $m_{i}$
reduces the influence of the transmission error with the fact that it reduces
the weight on the opinions
of others (which, especially in the long run, enables $i$ to aggregate signals
from all over the network and therefore provides very valuable information not
contained in $i$'s own signals). But he does take account of the fact that
when $m_{i}$ goes up, $y_{i}$ reflects more the information contained in $i$'s
signal as against what $i$ learnt from everyone else (which in the long run is
very close to what $i$'s neighbors \textit{too} learnt from everyone else) and
this is valuable for aggregate welfare. Technically, raising $m_{i}$
diminishes the correlation between $y_{i}$ and others' signals, and this
enhances the welfare of others.

We turn next to comparisons of the efficiency of information aggregation on
specific simple and oft-studied networks -- the complete network, the directed
circle and the star network. At the heart of our analysis is the
characterization of cumulated errors that each individual faces, and how then
each player mitigates the consequence of these errors by controlling the
weight of her own seed $x_{i}$ in his or her own long-run opinion, through the
choice of $m_{i}$. We find that the star network performs worse than the two
others, essentially because the central player propagates correlated errors to
all peripheral players, thus raising cumulated errors.


In Section 5, we use our example of the star network to address the key issue
of polarization. The result that $m_{i}$ is too low might suggest that there
is always too little disagreement in equilibrium. This is true for two-person
networks, but not in general. To see this consider a network where there are
two dense clusters (modeled as stars) connected by one link (say). Such a
network structure is not too dissimilar, for example, to the networks of
Republicans and Democrats in the US, who mostly communicate with each other
(Cox et al. (2020)). In this case, we show that lower $m_{i}$ is
associated with a high degree of consensus within each cluster but more
extreme polarization across the groups, reminiscent of the situation of the
Republicans and Democrats in the US. The general point, captured by Result 5,
is that social efficiency requires the dispersion of opinions \textit{within}
and \textit{between} subgroups to have same orders of magnitude.

Our very simple model, therefore, tells a useful story about why disagreements
are necessary, but also helps us understand why the resulting divergence of
opinions can be surprisingly large and when they are likely to be costly.

The rest of the paper is devoted to extensions. In Section 6, we allow for the
possibility idiosyncratic shocks in information transmission in addition to
permanent shocks. In this setting, the speed of updating, $\gamma_{i}$, which
plays no role in the previous analysis, also comes into play. Slowing down
updating by setting $\gamma_{i}$ close to zero allows the agent to minimize
the changes in opinions that result from these shocks, which is an advantage
because the shocks average out over time. This is what Result~6 shows.

In Section~7, we start by examining the implications of agents adding a slant
to the opinions they share--in other words adding errors that are biased in
some direction. Recent results from a survey experiment suggest this is a real
problem--people on social media are more likely to pass on messages that are
more concordant with their political opinions, somewhat irrespective of the
accuracy of the message \cite{pennycook21}. We note that biased errors do not
produce any essential changes in our analysis, though there is a further shift
towards reliance on one's own initial signal (higher $m$).

We next turn to the possibility of coarse communication--say each party only
reports their current best guess about which of two actions is preferable. In
this setting, the class of potentially \textquotedblleft natural" rules
include the infection models, studied in \cite{jackson08} among (many) others,
and the related class of models studied by \cite{ellison93,ellison95}, in
which agents may rely on the popularity of a particular action among
neighbors. We work with a version of this class of models where preferences
are heterogenous and each player has many neighbors. We show that systematic
errors in interpreting actions by neighbors makes the long-run outcome from a
DG-like rule entirely insensitive to the actual state of the world, but this
is not true for FJ-type rules. We use this framework to discuss the connection
between the errors we introduce and mis-specifications in Bayesian models (as
in \cite{frick20} and \cite{bohren21}) and the related (non-)robustness of
long-run beliefs.

To end Section~7 we highlight some examples where our findings \emph{are}
qualitatively altered. We have so far assumed that agents know the precision
of everyone's initial signals. We now explore the possibility that uncertainty
about the precision of everyone else's signal is the only source of friction
in communication. We find that, in the absence of transmission errors, this
\emph{does not} undermine the performance of DG-type rules. As a matter of
fact, in a set-up where each participant only knows the precision of their own
initial signal, perfect information aggregation can be achieved under DG, by
choosing $\gamma_{i}$ that is suitably scaled to the precision. This
observation delineates the key role played by transmission shocks in our
analysis, as opposed to other sources of shocks.

We next allow for the possibility that a friction comes from variations in who
speaks when. We show that under FJ rules long-run opinions are independent of
the communication protocol. In contrast, we show by example that the outcome
with DG rules is sensitive to who speaks when. So even in the absence of
noise, under protocol uncertainty, the performance of DG rules would be
impaired (though to a lesser extent than that induced by cumulated errors--
long-run opinions would \textit{not} blow up, but remain weighted averages of
initial opinions).

We conclude with a discussion of non-stationary rules and when and why they
may not always be appropriate.

\subsection{Related Literature}

Our paper contributes to the large literature on learning in social network
(see the excellent review by \cite{golub17}). We study non-Bayesian learning
on general networks with continuous choices and general networks. Within
Bayesian social learning, \cite{vives93,vives97} studies a setting similar to
ours (with agents receiving a noisy signal) and, unlike us, obtains long-run
convergence to the truth. The reason is that with continuous choice sets
Bayesian agents are able to perfectly extract the information content of the
noisy signals. When the choice set is coarser, aggregation can fail even with
Bayesian agents, as shown by \cite{banerjee92} or \cite{bikhchandani92}%
.\footnote{\cite{mossel15} shows that this result also depends on the network
structure and that for a large class of large networks, consensus and almost
perfect learning is possible even with coarse communication.}

In \cite{vives97}, like in this paper, agents underweight their private seed:
in his set up a stronger reliance on private signals \emph{in the initial
phase} would speed up learning and benefit all.\footnote{In the context of
non-Bayesian learning, \cite{mueller21} argue in related terms in favor of
non-stationary rules that aggregate information in a sufficiently dense part
of the network, before other agents get contaminated.} In our case, the weight
cannot be altered over time: however a higher reliance on private seeds
compared to equilibrium weights improves welfare because this limits both the
correlation between information sources and cumulated errors.

Our paper is also related to and inspired by the recent upsurge of interest in
the social learning with ``almost" Bayesian agents.
\cite{sethi12,sethi16,sethi19} allow for heterogenous and
\textit{unobservable} priors about the state, and since players exchange
beliefs (but not priors), there can be long-run disagreement. However the
divergence cannot exceed the spread in initial biases because agents interpret
correctly the reports of others based on the known distribution of priors. In
contrast, \cite{eyster10}, \cite{frick20}, \cite{bohren21} and
\cite{gentzkow21}, among others, introduce mis-specifications that lead agents
to \textit{incorrectly interpret} reports or actions of others.
In \cite{eyster10}, the errors are assumed to be significant enough to
generate incorrect long-run beliefs for many signal realizations. By contrast,
\cite{frick20} show that even small systematic mis-specifications can lead to
interpretation errors that cumulate over time, though, as shown in
\cite{bohren21}, a restriction to a small number states and common priors can
prevent this drift (See Section~\ref{subsectionMisspecified} for an extended
discussion of the connection between these two papers and ours). Finally, in
\cite{gentzkow21}, uncertain precision of signals and mis-specifications lead
players to overestimate the precision of signals received by others who are
similarly biased,

Other papers directly modify the updating rule itself. \cite{jadbabaie12}
introduce rules that combine Bayesian updating of own signals with a DG-like
averaging over neighbors' beliefs, while \cite{levy15} consider a rule which
involves cumulating log likelihood-ratios, which they justify, like DG, on the
ground that it mimics what a subjective Bayesian (with an erroneous model of
the world) would do (see also \cite{dasaratha23}. Finally \cite{molavi18}
provides axiomatic justification(s) (motivated by imperfect recall) for DG
style linear aggregation (and averaging) of log
belief-ratios.\footnote{Attempts to provide axiomatic foundations of the DG
rule in the statistics literature go back to \cite{genest86}.}

By contrast we take an evolutionary approach to rule selection, assuming
selection within a \textit{restricted} family of plausible stationary rules.
There is of course a vast literature on the evolutionary selection of general
behavioral rules, going back to \cite{axelrod84}. \cite{fudenberg98} provide
an excellent introduction to the selection of strategies in game theoretic
settings. Our focus is on selecting
rules for aggregating information in potentially large and complex network
settings.

\section{Basic Model}

\subsection{Transmission on the network}

We consider a finite network with $n$ agents, assume noisy
transmission/reception of information and define a simple class of rules that
players may use to update their opinions.

Formally, each agent $i$ in the network has an \textit{initial opinion}
$x_{i}$ and, at date $t$, \textit{an opinion} $y_{i}^{t}$ that can both be
represented as real numbers.\footnote{This opinion can be interpreted as a
point-belief about some underlying state, which will eventually be used to
undertake an action.} Taking as given the matrix $A$ characterizing the
weights $A_{ij}$ that $i$ puts on $j$'s opinion, we consider the class of
updating rules (FJ) parameterized by the weights $m_{i}$ and $\gamma_{i}$ and
specified in the introduction. Along with the Expression~(\ref{eqz}) for
transmission errors, the dynamic of opinions for player $i$ is:
\[
y_{i}^{t}=(1-\gamma_{i})y_{i}^{t-1}+\gamma_{i}(m_{i}x_{i}+(1-m_{i})(\sum_{j\in
N_{i}}A_{ij}y_{j}^{t-1}+\varepsilon_{i}^{t})
\]

When $m_{i}=0$, the rule corresponds to the well-studied DeGroot rule (DG).
When $m_{i}>0$, then in each period the rule mixes decision-maker's own
initial opinion $x_{i}$ with DG. This perpetual use of the initial opinion in
the updating process gives FJ a non-Bayesian flavor, since for a Bayesian,
their prior (i.e., the seed) is already integrated into $y_{i}^{t-1}$ and
therefore there is no reason to go back to it.\footnote{In fact, as mentioned
already, the one obvious attraction of $DG$ is its quasi-Bayesian flavor. Note
that although formally the expression (\ref{FJ}) encompasses the DG rule, we
shall refer to FJ as a rule for which $m_{i}>0$.}

To avoid technical difficulties once we give agents discretion in choosing
their updating rule, we set $\underline{\gamma}>0$ arbitrarily small and
restrict attention to FJ rules where $\gamma_{i}\geq\underline{\gamma}$. We
also assume that the matrix $A$ is \textit{connected} in the sense that for
some positive integer $k$, the $k^{th}\ $power of $A$ only has strictly
positive elements,\textbf{ }i.e., $A_{ij}^{k}>0$ for all $i,j.$ In other words
everyone is within a finite number of steps of the rest.

Note that all the rules considered here are stationary, in the sense that the
weighting parameters $m_{i}$ and $\gamma_{i}$ do not vary over
time.\footnote{In this sense even DG is only quasi-Bayesian, since for
Bayesian the weight on new reports goes down over time.} We see these as
plausible ways in which boundedly rational agents might incorporate others'
opinions into their current opinion. We recognize that with enough knowledge
of the structure of the network and the process by which new information gets
incorporated, adjusting the weights over time may make sense and return to
this possibility in Section \ref{SectionNonStationary}.

We also impose the assumption that everyone operates on the same time
schedule: periods are defined so that everyone changes their opinion once
every period and everyone else get to observe that change of opinion before
they adjust their opinion in the following period. We will discuss what
happens if we relax this assumption in Section \ref{SectionProtocols}.

\subsection{Errors in opinion sharing}

The term $\varepsilon_{i}^{t}$ is an important ingredient of our model, meant
to capture some imperfection in transmission.\footnote{There has been several
recent attempts to introduce noisy or biased transmission in networks. In
\cite{jackson19}, information is coarse (0 or 1), and noise can either induce
a mutation of the signal (from 0 to 1 or 1 to 0) or a break in the chain of
transmission (information does not get communicated to the network
neighbor).}
It represents a distortion in what each individual \textquotedblleft
hears\textquotedblright\ that aggregates all the different sources of errors.
Until Section 6, we assume that the error term is persistent, realized at the
start of the process and applying for the duration of the updating
process.\footnote{One interpretation is that each information aggregation
problem is characterized by the realization of an initial opinion vector $x$
and persistent bias vector $\xi$, and that agents face a distribution over
problems.} We shall denote by $\xi_{i}$ this persistent error, so
\[
\varepsilon_{i}^{t}\equiv\xi_{i}%
\]
In Section 6, we extend the model and incorporate idiosyncratic errors:%
\[
\varepsilon_{i}^{t}=\xi_{i}+\nu_{i}^{t},
\]
where $\nu_{i}^{t}$ are i.i.d. across time and agents.

We interpret $\xi_{i}$ as a systematic bias that slants how opinions of others
are \textit{processed} by $i$. Biases $\xi_{i}$ may be drawn independently
across players, but we shall also discuss cases where they are positively
correlated, such as when a group of friends share a political bias. Also note
that although errors are indexed by $i$, our formulation can accommodate
biases that result from both\textit{ }\textquotedblleft hearing" errors and
\textquotedblleft sending" errors.\footnote{For example, if there were both
\textquotedblleft hearing" errors labelled $\xi_{i}^{h}$ and \textquotedblleft
sending" errors labelled $\xi_{i}^{s},$ one could define $\xi_{i}=\xi_{i}%
^{h}+\sum_{j}A_{ij}\xi_{j}^{s}$ as the resulting processing error. Sending
errors naturally generate correlations across the $\xi_{i}$'s, and a profile
of errors that depend on the network structure $A$. This is further discussed
in the Appendix.}

For convenience, we assume that all error terms are unbiased (that is,
$E\xi_{i}=0$ and $E\nu_{i}^{t}=0$)
and homogenous across players, so we let
\[
\varpi=\varpi_{i}=var(\xi_{i})
\]

\subsection{The objective function}

There is an underlying state $\theta$, and agents want their decision to be as
close as possible to that underlying state, where the decision is normalized
to be the same as the agent's long-run opinion. In other words, we visualize a
process where agents exchange opinions a large number of times before the
decision needs to be taken.

Given this private objective, we explore each agent's incentives to choose his
updating rule within the class of FJ rules to maximize the above objective on
average across many different realizations of the underlying state of the
world, the initial opinions and the transmission errors. We have in mind the
idea that individuals choose a single rule to apply to many different
problems. This is why we focus on their ex ante performance.\footnote{That is,
on average over states, initial opinions and transmission errors.} The set of
possible updating rules is extraordinary vast, so the limitation to FJ rules
is of course a restriction. Our motivation is to examine the incentives of
\textit{mildly} sophisticated agents who have some limited discretion over how
they update opinions.\smallskip{}

Formally, we assume that the initial signals are given by
\[
x_{i}=\theta+\delta_{i}%
\]
where the $\theta$ are drawn from some distribution $G(\theta)$ with mean zero
and finite variance, $\delta_{i}$, $\xi_{i}$ and $\nu_{it}$ are random
variables that are independent of each other for all $i$ and $t$ and are also
independent of $\theta.$ We assume that noise terms $\delta_{i}$ are unbiased,
with variance $\sigma_{i}^{2}>0$. For convenience, except where we need to
assume otherwise to make a specific point, we set $\sigma_{i}=1$ for all $i$,
but we do not actually need this assumption.

For any $t$, each profile of updating rules $(m,\gamma)$ generates at any date
$t$, a distribution over date $t$ opinions. We now define the expected loss
(where the expectation is taken across realizations of $\theta$, $\delta_{i}$,
and $\varepsilon_{i}^{t}$ for all $i$ and $t$):
\[
L_{i}^{t}=E(y_{i}^{t}-\theta)^{2}%
\]
We then define the limit loss $L_{i}=\lim_{t\nearrow\infty}L_{i}^{t}%
$.\footnote{Alternatively, one could define $L_{i}=\lim_{h\searrow0}(1-h)\sum
h^{t-1}L_{i}^{t}$, assuming that the agent makes a decision at a random date
far away in the future and that his preference over decisions is $u_{i}%
(a_{i},\theta)=-(a_{i}-\theta)^{2}$.
\par
$L_{i}$ is well-defined for any vector $m,\gamma$ so long as $m\neq0$. As it
will turn out, for $m=0$, $L_{i}$ is infinite. Note that each player can
secure $L_{i}\leq var(\delta_{i})=\sigma_{i}^{2}=1$ by ignoring everyone
else's opinions ($m_{i}=1$).}


\subsection{Methodological assumptions\label{SectionMA}}

The loss $L_{i}$ depends on the profile of updating rules $(m,\gamma)$, and
our main methodological assumptions are that (i) there is a force towards the
use of higher performing rules (e.g., justified by evolution or reinforcement
learning), and (ii) in this quest for higher performing rules, each individual
considers (and gets feedback about) only a limited set of rules (i.e., the FJ class).

Formally, our analysis boils down to examining a rule-choice game where, given
the rules adopted by others, each agent aims at minimizing $L_{i}$ (using the
instruments $m_{i}$ and $\gamma_{i}$ available to her): the object of interest
is the Nash equilibrium of this rule choice game. Since $L_{i}$ is an
expectation across various realizations of initial signals and noise in
transmission, we think of the person choosing one rule, parameterized by
$m_{i}$ and $\gamma_{i}$, to apply in many different life situations. These
parameters are meant to capture some general features of opinion formation:
specifically the \textit{persistence} of initial opinions, and \textit{speed
of adjustment} of the current opinion.\footnote{Our view is that these
features probably do adjust to the broad economic environment agents face, but
for each opinion-formation problem within a certain context, the actual
sequence of opinions is mechanically generated given these features.}

It is precisely this fact that
rules apply across many different problems\textbf{,} and that a limited set of
rules are considered, that makes our third route cognitively less demanding
than the Bayesian route. While we agree that choosing $m_{i}$ and $\gamma_{i}$
optimally is a difficult problem which in principle requires knowledge of the
structure of the model, there is no reason why the standard justification of
Nash Equilibrium as a resting point of an (un-modeled) learning/evolutionary
process would not apply here. Moreover, one of our most important results is
that DG rules, and indeed all rules that put too little weight ($m_{i}$) on
initial opinions, are dominated when there is noise in transmission,
suggesting a strong force away from DG even if agents find it difficult to
find the exact optimal value of $m_{i}$.

In the next Section we start by exploring the long-run properties of different
learning rules within the DG and FJ class, with and without errors. Then we
turn to the optimal choice of learning rules.


\section{Some properties of the long-run opinions}

In the paper we make a distinction between Results, which are meant to be of
substantive interest, and Propositions, which are more technical and are meant
to explain and lead up to the Results. This section reports a number of
Propositions that provide the bulwark for our main results in Section~4. We
start by studying the properties of long-run opinions under DG and FJ with and
without errors. In particular, we shall show that in the presence of errors
there is convergence under FJ as long as at least one person $i_{0}$ has
$m_{i_{0}}>0$, but not under DG. We then explore what determines the variance
of the limit opinion in the case where such a limit opinion exists. In
particular what part of it comes from the \textquotedblleft
signal\textquotedblright-- the original seeds -- and what part from the noise
that gets added along the way? We also explore the degree to which a player
can influence long-run opinions through the choice of $m_{i}$ and $\gamma_{i}$.

\subsection{DG without errors.\label{sectionDGwithout}}

It is well-known that in the DG case without errors ($m_{i}=0$ for all $i$)
learning converges to consensus and steady state values of $y_{i}$ for all
${i}$. Define $\Gamma$ as the diagonal matrix such that $\Gamma_{ii}%
=\gamma_{i}$. In matrix form, the dynamic of the vector of opinions
$y^{t}=(y_{i}^{t})_{i}$ under DG without noise can be expressed as%
\begin{equation}
y^{t}=B_{0}y^{t-1}\text{ where }B_{0}=I-\Gamma+\Gamma A, \label{EqDG0}%
\end{equation}
implying that
\begin{equation}
y^{t}=(B_{0})^{t}x \label{EqDG1}%
\end{equation}
where $x$ is the vector of initial opinions. Let $\Delta_{n}$ be the set of
vectors of non-negative weights $p=\{p_{i}\}_{i}$ with $\sum p_{i}=1$. Because
the network is connected, $A$ is a irreducible stochastic
matrix,\footnote{This is because $A^{k}$ only has strictly positive elements
for some large $k$.} so there is a (unique) strictly positive vector of
weights $\rho\in\Delta_{n}$ such that $\rho A=\rho$. When $\gamma_{i}>0$ for
all $i$, $B_{0}$ is also an irreducible stochastic matrix, so there is a
unique vector $\pi\in\Delta_{n}$ such that $\pi B_{0}=\pi$, and we must
have\footnote{This is because $\pi^{0}\equiv\rho\Gamma^{-1}$ solves $\pi
^{0}B_{0}=\pi^{0}-\rho+\rho=\pi^{0}$. Thus, since\ $\pi$ is unique, $\pi$ must
be proportional to $\pi^{0}$.}%
\begin{equation}
\frac{\pi_{i}}{\pi_{j}}\equiv\frac{\rho_{i}}{\rho_{j}}\frac{\gamma_{j}}%
{\gamma_{i}} \label{pi}%
\end{equation}
When $t$ gets large, all rows of $(B_{0})^{t}$ converge to $\pi$, so all
opinions $y_{i}^{t}$ converge to the same limit opinion $\pi.x$, i.e.,
\begin{equation}
y_{i}=\pi.x\text{ for all }i\text{.} \label{yi}%
\end{equation}
So although the direct contribution of $i$'s initial signal to $i$'s opinion
vanishes, it surfaces back from the influence of neighbors' opinions (which
increasingly incorporate $i$'s initial signal), settling at a limit weight
equal to $\pi_{i}$.

Using (\ref{pi}), one may rewrite (\ref{yi}) to highlight how the speed of
adjustment $\gamma_{i}$ affects player $i$'s influence on long-run opinions.
We have:\medskip

\textbf{Proposition 0:} \textit{When }$m_{i}=0$\textit{ for all }$i$\textit{
and in the absence of errors, long-run opinions all converge to the same limit
opinion }$\pi.x$\textit{ and}
\begin{equation}
y_{i}=\pi_{i}x_{i}+(1-\pi_{i})q^{i}.x_{-i}\text{ \textit{where} }\frac{\pi
_{i}}{1-\pi_{i}}=\frac{1}{\gamma_{i}}\frac{\rho_{i}}{\sum_{j\neq i}\rho
_{j}/\gamma_{j}} \label{no error}%
\end{equation}
\textit{and where }$q^{i}$\textit{ is a probability vector in }$\Delta_{n-1}%
$\textit{ that does not depend }$\gamma_{i}$\textit{. }\medskip

In other words, the network structure determines $\rho$. Given $\rho$, player
$i$ can use $\gamma_{i}$ to control her influence on the long run opinion,
$\pi_{i}$, but she cannot control the relative weights on the opinions of
others, captured by $q^{i}$.

\subsection{DG with errors: exploding dynamics.}

We show below that if all agents follow a DG rule, then for almost all
realization of $\xi$, the long-run opinions diverge. \medskip

\textbf{Proposition 1}. \textit{Assume that }$m_{i}=0$ \textit{for all }%
$i$\textit{. Then }for almost all\textit{ realizations of }$\xi$,\textit{
}$\lim\left\vert y_{i}^{t}\right\vert =\infty$\textit{ for all }$i$ and $x$.
\medskip

This proposition shows, for one, that an error $\xi_{1}$ in a single agent's
perception is enough to drive everyone's opinions arbitrarily far from the
truth: if $\xi_{1}>0$, say, the error creates a discrepancy between $1$'s
opinion and that of the others, but every time the others' opinions catch up
with him, agent 1 further raises his opinion compared to others, prompting
another round of catching up, and eventually all opinions blow up.

\textbf{Proof:} With errors, Equations \ref{EqDG0} and \ref{EqDG1} become
$y^{t}=B_{0}y^{t-1}+\Gamma\xi$ and%
\[
y^{t}=(B_{0})^{t}x+\sum_{0\leq k<t}(B_{0})^{k}\Gamma\xi
\]
For $k$ large enough, each row of $(B_{0})^{k}$ is close to $\pi$, so
$y_{i}^{t}$ diverges for all $i$ whenever $\pi\Gamma\xi\neq0$.$\blacksquare$

\subsection{Anchored dynamics under FJ.}

Fixing again $x$ and $\xi$, we now examine long-run dynamics under FJ.\medskip

\textbf{Proposition 2.} \textit{Assume at least one player, say }$i_{0}%
$,\textit{ updates according to FJ (with }$m_{i_{0}}>0$\textit{).Then, for any
fixed }$x$ \textit{and} $\xi$\textit{, }$y^{t}$\textit{ converges, and the
limit vector of opinions }$y$\textit{ does not depend on }$\gamma$
\textit{nor} \textit{on the signal }$x_{i}$\textit{ of any individual with
}$m_{i}=0$\textbf{. }\medskip

Proposition 2 shows that to prevent all the opinions from drifting away, it is
enough that there is one player who continues to put at least a minimum amount
of weight on his own initial opinion in forming his opinion in every period.
Proposition 2 also shows that when $m_{i}=0$, the signal initially received by
$i$ has no influence on the players' long-run opinions. A detailed proof is in
the Appendix.\smallskip{}

When $m_{i_{0}}>0$ for some $i_{0}$, proving convergence is
standard.\footnote{The argument follows \cite{friedkin99}.} The limit opinion
$y$ then solves
\[
y_{i}=(1-\gamma_{i})y_{i}+\gamma_{i}(X_{i}+(1-m_{i})A_{i}y)\text{ for all }i.
\]
where $X_{i}=m_{i}x_{i}+(1-m_{i})\xi_{i}$, which implies that, in matrix form,
it is also the solution of\textit{ }
\begin{equation}
y=X+(I-M)Ay \label{eqFJ0}%
\end{equation}
where $M$ is the diagonal matrix with $M_{ii}=m_{i}$. This expression implies
that limit opinions are independent of the $\gamma_{i}$'s. It also explains
why long-run opinions only involves the seeds $x_{i}$ of players for whom
$m_{i}>0$, since for the others, $X_{i}=\xi_{i}$.

\subsection{The dominance of noise under low $m$.}

Although convergence is guaranteed when at least one player does not use DG,
there is no discontinuity at the limit where \textit{all} $m_{i}$ get small:
long-run opinions then become highly sensitive to the persistent error $\xi$.
We have:\medskip

\textbf{Proposition 3}: Let $\overline{m}=\max m_{i}.$ Then $L_{i}\geq
\frac{\varpi}{n}\frac{(1-\overline{m})^{2}}{\overline{m}^{2}}$.\medskip

The detailed proof is in the Appendix. The lower bound on $L_{i}$ is obtained
by showing that for given $x,\xi$, long-run expected opinions are a weighted
average of \textit{modified initial opinions}, defined, whenever $m_{i}>0$,
as
\[
\widetilde{x}_{i}=x_{i}+(1-m_{i})\xi_{i}/m_{i}.
\]
To fix ideas, assume $m_{i}>0$ for all $i$.\footnote{The argument generalizes
to the case where a subset $N^{0}\varsubsetneq N$ of agents follows DG
($m_{i}=0$).
(see Appendix).} Then one can write (using the previous notation)
$X=M\widetilde{x}$ and obtain, using (\ref{eqFJ0})%
\begin{equation}
y=M\widetilde{x}+(I-M)Ay\equiv P\widetilde{x} \label{Av1}%
\end{equation}
where $P$ is a probability matrix.\footnote{This means that each line of $P$
is a probability vector. $P$ is the limit of $P^{t}$ defined recursively by
$P^{t+1}=M+(I-M)AP^{t}$ and $P^{1}=I$. By induction, each $P^{t}$ (and $P)$ is
a probability matrix.}
Intuitively, $x_{i}$ can be thought of the \textit{seed} that individual $i$
plants in her belief in every period, and $\widetilde{x}_{i}$ as the
\textit{effective seed} given processing errors. Long-run opinions are
averages over effective seeds. Since the variance of each $\widetilde{x}_{i}$
is bounded below by $\frac{\varpi(1-\overline{m})^{2}}{\overline{m}^{2}}$, we
obtain the desired lower bound.\medskip{}

\textit{The two-player case.} The two-player case provides a useful
illustration. With two players, assuming $m_{1}$ and $m_{2}$ strictly
positive, long-run opinions solve%
\[
y_{i}=m_{i}\widetilde{x}_{i}+(1-m_{i})y_{j}=m_{i}\widetilde{x}_{i}%
+(1-m_{i})(m_{j}\widetilde{x}_{j}+(1-m_{j})y_{i}%
\]
which further implies%
\begin{equation}
y_{i}=p_{i}\widetilde{x}_{i}+(1-p_{i})\widetilde{x}_{j}\text{ where }%
p_{i}=\frac{m_{i}}{m_{i}+(1-m_{i})m_{j}} \label{eq2}%
\end{equation}
confirming that long-run opinions are weighted average of modified opinions.
Furthermore%
\begin{equation}
y_{i}=p_{i}x_{i}+(1-p_{i})(x_{j}+\widehat{\xi}_{i})\text{ where }\widehat{\xi
}_{i}=\frac{\xi_{i}+\xi_{j}}{m_{j}}-\xi_{j}. \label{eq2b}%
\end{equation}
The term $\widehat{\xi}_{i}$ can be interpreted as the \textit{cumulated
error} that player $i$ faces, resulting from each player repeatedly processing
the other's opinion with an error, while $p_{i}$ characterizes how player
$i$'s own seed \textit{influences} her long-run opinion. Since $p_{i}%
+p_{j}=\frac{m_{1}+m_{2}}{m_{1}+m_{2}-m_{1}m_{2}}>1$, it must be that players
differ in the weight they each put in the long-run on their seeds, so there is
disagreement, and the magnitude of the disagreements rises with $m$.

In networks, \textit{echo effects} arise because players incorporate opinions
that they have themselves contributed to shape, and these echoes shape both
\textit{long-run influence }and \textit{cumulated errors}: when $m_{i}$ is
small, the influence of player $i$ may nevertheless be large because although
$i$ puts a large weight on $y_{j}$, if $m_{j}/m_{i}$ is small as well then
$y_{j}$ has been mostly shaped by $x_{i}$; echoes also shape cumulated errors
because a single loop of communication generates a combined error of $\xi
_{i}+\xi_{j}$, which is (partially -- but almost entirely when $m_{j}$ is
small) added to all opinions and thus cumulates over time.

\subsection{Influence under FJ rules and cumulated errors.}

Under DG rules and no errors, a player can control her influence by modifying
$\gamma_{i}$. Under FJ rules, the long-run opinions do not depend on
$\gamma_{i}$--instead, as the previous two-player example illustrates, the
limit opinions depend on the vector of weights $m$. Here we characterize both
influence and cumulated errors for more general networks.

When at least one player $i_{0}$ sets $m_{i_{0}}>0$, long-run opinions
converge and we have%

\begin{equation}
y_{i}=m_{i}x_{i}+(1-m_{i})\xi_{i}+(1-m_{i})\widehat{y}_{i}\text{ with
}\widehat{y}_{i}\equiv\sum_{k\neq i}A_{ik}y_{k} \label{Eqyi}%
\end{equation}
Player $i$'s opinion thus builds on the opinion $\widehat{y}_{i}$ of a
(fictitious)\textbf{ }\textit{composite neighbor} who aggregates the
opinions\textbf{ }$y_{k}$\textbf{, }to which the error $\xi_{i}$ is added.
Letting $\widetilde{A}_{kj}^{i}=\frac{A_{kj}}{1-A_{ki}}$, we rewrite
(\ref{Eqyi}) to describe how each opinion $y_{k}$ builds on $y_{i}$:
\begin{equation}
y_{k}=m_{k}x_{k}+(1-m_{k})\xi_{k}+(1-m_{k})A_{ki}y_{i}+(1-m_{k})(1-A_{ki}%
)\sum_{j\neq k,i}\widetilde{A}_{kj}^{i}y_{j} \label{Eqyk}%
\end{equation}
So in effect, in incorporating the composite opinion $\widehat{y}_{i}$, player
$i$ is (partially) incorporating her own opinion $y_{i}$: the opinions that
$i$ gets from others are partially echoes of her own opinion. So even if her
per-period reliance on $x_{i}$ is small (i.e. $m_{i}$ small), her seed $x_{i}$
may eventually have a large influence on long-run opinions. Another aspect in
that in incorporating the composite opinion $\widehat{y}_{i}$, each player $i$
is (partially) adding other players' error terms to her own, and any opinion
that contributes to $\widehat{y}_{i}$ is itself subject to errors.
Proposition~4 below characterizes both effects: long-run influence and
cumulated errors.

Let $M^{i}$\ (resp. $\alpha^{i}$) be the diagonal $N-1$\ matrix for which
$M_{kk}^{i}=m_{k}$\ for $k\neq i$\ (resp. $\alpha_{kk}^{i}=A_{ki}$) and define
the matrix $Q^{i}=(I-(I-M^{i})(I-\alpha^{i})\widetilde{A}^{i})^{-1}$\ and
vector $R^{i}$\ such that $R_{j}^{i}=\sum_{k}A_{ik}Q_{kj}^{i}$. Also let
$h_{i}\equiv1/\sum_{j\neq i}R_{j}^{i}m_{j}$. We have:\smallskip\smallskip

\textbf{Proposition 4:} \textit{ Assume player }$i_{0}\neq i$ has\textit{
}$m_{i_{0}}>0$. \textit{Then }$h_{i}\geq1$ \textit{and }%
\begin{align}
y_{i}  &  =p_{i}x_{i}+(1-p_{i})(\widehat{x}_{i}+\widehat{\xi}_{i})\text{
\textit{where }}\widehat{x}_{i}=q^{i}.x_{-i},\label{Propyi}\\
\frac{p_{i}}{1-p_{i}}  &  =\frac{m_{i}h_{i}}{(1-m_{i})}\text{, }q_{j}%
^{i}=\frac{R_{j}^{i}m_{j}}{\sum_{j\neq i}R_{j}^{i}m_{j}}\text{ and
}\nonumber\\
\text{ }\widehat{\xi}_{i}  &  =h_{i}(\xi_{i}+\sum_{j\neq i}R_{j}^{i}\xi
_{j}(1-m_{j}))\nonumber
\end{align}

Proposition 4 provides an analog of Proposition 0 when at least one player
uses an FJ rule. Without errors, player$~i$'s long-run opinion is an average
between her own seed $x_{i}$ and a \textit{composite seed} $\widehat{x}_{i}$
(an average over the others' seeds). The weight $p_{i}$ defines how player
$i$'s own seed influences her long-run opinion, and through the choice of
$m_{i}$ player$~i$ has full control over this weight. Player$~i$ however has
no control over the composite seed $\widehat{x}_{i}$, as the vector of weights
$q^{i}\in\Delta_{n-1}$ is fully determined by $A$ and $m_{-i}$.

In the presence of errors, the weights $p_{i}$ and $q^{i}$ remain the same.
The difference is that when attempting to incorporate the composite seeds,
player$~i$ faces a cumulated error term $\widehat{\xi}_{i}$. This error term
can be very large when all $m_{j}$ are small.

Proposition 4 also confirms an insight suggested by Proposition 2: the seed
$x_{j}$ of any individual that sets $m_{j}=0$ has no influence on long-run
opinion (either own or others). Finally, to complete the set of possible
cases, we have:\medskip

\textbf{Proposition~5}: \textit{If }$m_{-i}=0$\textit{ and }$m_{i}>0$\textit{,
then }$y_{i}=x_{i}+\frac{1-m_{i}}{m_{i}}(\xi_{i}+\sum_{j\neq i}R_{j}^{i}%
\xi_{j})$\textit{ where }$R$\textit{ is as defined in Proposition 4.}\medskip

Consistent with Proposition 3, echo effects rise without bound when $m_{i}$
gets small.
Proposition~4 and 5 imply that if all players but $i$ use DG, all players
opinion's will build on $x_{i}$ only, however small $m_{i}$ is.
\cite{mueller17} makes a similar observation in a model without errors
(concluding that learning outcomes are highly sensitive to small departures
from $DG$).\medskip

We now use Proposition~4 to provide a characterization of the privately
optimal choice of $m_{i}$, and its consequence for the loss $L_{i}$.
Recall from Proposition 4 that $y_{i}=p_{i}x_{i}+(1-p_{i})(\widehat{x}%
_{i}+\widehat{\xi}_{i})$ where $\widehat{x}_{i}+\widehat{\xi}_{i}$ is a term
that only \textit{depends on the structure of the network and }$m_{-i}$, and
which has variance
\begin{equation}
W_{i}\equiv var(\widehat{x}_{i})+\widehat{\varpi}_{i}\text{ where
}\widehat{\varpi}_{i}=E\widehat{\xi}_{i}^{2} \label{W}%
\end{equation}
Since player $i$ fully controls $p_{i}$ by adjusting $m_{i}$ (since
$\frac{p_{i}}{1-p_{i}}=\frac{h_{i}m_{i}}{1-m_{i}})$, individual $i$ optimally
sets $p_{i}$ so that $\frac{p_{i}}{1-p_{i}}=W_{i}/\sigma_{i}^{2}$, and we
obtain:\medskip

\textbf{Proposition 6:} \textit{For a given }$m_{-i}$\textit{, the optimal
choice of }$m_{i}$\textit{ and resulting loss }$L_{i}$ \textit{satisfy}%
\begin{equation}
\frac{m_{i}}{1-m_{i}}=\frac{W_{i}}{h_{i}\sigma_{i}^{2}}\text{ \quad
\textit{and}\quad\ }L_{i}=\sigma_{i}^{2}p_{i}=\frac{W_{i}}{1+W_{i}/\sigma
_{i}^{2}}. \label{LP6}%
\end{equation}


Since $W_{i}/h_{i}$ depends on the network structure and $m_{-i}$ only,
Proposition~6 will allow us to easily characterize equilibrium weights
$m_{i}^{\ast}$, as well as the induced equilibrium losses.

This Proposition also implies that the loss $L_{i}$ is fully determined by
$W_{i}$. It shall be instructive to compare $L_{i}$ with the minimum feasible
loss $v^{\ast}$ obtained under efficient aggregation of initial opinions,
i.e., $v^{\ast}=\min_{q}var(\pi.x)$. This minimum loss satisfies:
\begin{equation}
v^{\ast}=\sigma_{i}^{2}\pi_{i}^{\ast}=\frac{\underline{W}_{i}^{\ast}%
}{1+\underline{W}_{i}^{\ast}/\sigma_{i}^{2}} \label{LP6b}%
\end{equation}
where $\underline{W}_{i}^{\ast}=\min_{q}var(q.x_{-i})$. \footnote{This is
because $v^{\ast}=\min_{\pi}\pi.x=\min_{\pi_{i}}var(\pi_{i}x_{i}+(1-\pi
_{i})\underline{W}_{i}^{\ast})$.} So whenever $W_{i}$ rises above
$\underline{W}_{i}^{\ast}$, the loss $L_{i}$ rises above $v^{\ast}$.
Expression (\ref{W}) thus highlights the two possible additional sources of
losses that player $i$ now faces: (i) the fact that seeds of others may not be
efficiently aggregated (i.e. $var(\widehat{x}_{i})>\underline{W}_{i}^{\ast}$)
and (ii) the presence of the cumulated error term $\widehat{\xi}_{i}$.

Section 4 will build upon Propositions~4,~5 and 6 to characterize the
equilibrium of the rule choice game. We shall also see that when errors are
small, the cumulated errors $\widehat{\xi}_{i}$ are the preponderant source of
inefficiency.
%
We conclude this Section with further comments on DG and FJ rules.

\subsection{Understanding the difference between DG and FJ\label{comments}}

(a) \textbf{On anchoring, influence and consensus:} DG and FJ generate a very
different dynamic of opinions. Permanently putting weight on one's initial
opinion is equivalent to putting a weight on the opinion of an individual that
never changes opinion: it anchors one's opinion, preventing too much drift. As
a result, it also anchors the opinions of one's neighbors, hence, the opinions
of everyone in the (connected) network.

The channel through which each player influences long-run opinions also
differs substantially. In the absence of noise, and for a given network
structure, relative influence in DG depends on relative speed of adjustment
$\gamma$, with lower speed increasing influence (see (\ref{pi})).


In contrast, under $FJ$, the speeds of adjustment $\gamma$ have no effect on
long-run opinions $y$. Only the $m_{i}$'s (and the structure of the network)
matter. These $m_{i}$'s determine \textit{player-specific} vectors of weights,
but at the limit where all $m_{i}$'s are very small, these vectors converge to
one another\textbf{ }(see Appendix), with the weight $p_{i}$ on $i$'s seed
proportional to $m_{i}\rho_{i}$, that is:
%

\begin{equation}
\frac{p_{i}}{p_{k}}=\frac{m_{i}\rho_{i}}{m_{k}\rho_{j}} \label{pioverpk}%
\end{equation}
This is an analog to (\ref{pi}) showing that close to the limit, $m_{i}$ plays
the same role as $1/\gamma_{i}$ does in DG and consensus obtains. As the
$m_{i}$'s go up however, consensus disappears: players \textquotedblleft agree
to disagree".\textbf{\bigskip}

(b) \textbf{On the fragility of DG: } There is something inherently fragile
about the long-run evolution of opinions under DG. Since individuals don't put
any weight on their own initial signal after the first period, the direct
route for that signal to stay relevant is through the weight put on their own
previous period's opinion. This source clearly has dwindling importance over
time. This gets compensated by the growing weight on the indirect route--each
individual $i$ adjusts his or her opinion based on the opinions of their
neighbors, and these are in turn influenced by $i$'s past opinions and through
those, by $i$'s initial signal. In DG without transmission errors, the second
force at least partly
offsets the first one\textbf{~}--\textbf{~}but this is no longer true when
there is any transmission error because of the cumulative effect of noise that
comes with the feedback from others.\bigskip{}

(c) \textbf{On the source of change in opinion}: One way to assess the
difference between DG and FJ is to express them in terms of changes of
opinions and opinion spreads. Defining the change of opinion $Y_{i}^{t}%
=y_{i}^{t}-y_{i}^{t-1}$, the neighbors' average opinion $\widehat{y}_{i}^{t}$
and the spread $D_{i}^{t}=\widehat{y}_{i}^{t}-y_{i}^{t}$ between others' and
own opinions, and setting $\gamma_{i}=1$ for all $i$ for the FJ process, we
have the following expressions:
\begin{align}
Y_{i}^{t}  &  =\gamma_{i}(D_{i}^{t-1}+\xi_{i})\text{ }\tag{DG}\\
Y_{i}^{t}  &  =(1-m_{i})A_{i}Y^{t-1} \tag{FJ}%
\end{align}
Under DG, one changes one's opinion whenever there is a (perceived) difference
between that opinion and the opinions of one's neighbors: any difference
generates an adjustment aimed at reducing it. In the absence of errors, this
creates a force towards consensus, with $D_{i}^{t}$ and $Y_{i}^{t}$ eventually
converging to $0$. With errors however, this adjustment aimed at reducing the
(perceived) spread actually keeps opinions moving:\footnote{Technically,
opinions can never settle because this would require finding a vector $y$ for
which $D+\xi=0,$ hence $Ay-y+\xi=0$ which is not possible unless $\rho.\xi
=0$.} errors are eventually incorporated into the opinions of all the players,
and repeated errors tend to cumulate and generate a general drift in opinions.
The force towards consensus is in this sense too strong.

By contrast, under FJ, players only incorporate \textit{changes }in the
opinions of others\textbf{.} So, in the case where the transmission error is
fixed, $\xi_{1}$ will generate a \textit{one time }change on $1$'s opinion,
but it won't, by itself, generate any further changes for player 1. Of course,
this initial (unwanted) change of opinion will trigger a sequence of further
changes -- it will be partially incorporated in player 2's opinion, and
therefore come back to player 1 again. This is what we call an \textit{echo
effect}. But, when $m_{i}>0$ for at least one player, the echo effect will be
smaller than the initial impact and will get even smaller over time, and as
result, opinions won't blow up: all $Y_{i}^{t}$'s eventually converge to $0$.
Nevertheless, if all $m_{i}$ are small, the echo effects are not dampened
enough, and the consequence is a high sensitivity of the final opinion to the errors.

\section{Choosing the rule}

\subsection{When there are no errors}

We build upon Proposition~4 and 5 to characterize the equilibrium of the
rule-choice game, starting with the case of no error. We show that the
equilibrium must be DG and that in equilibrium, information aggregation must
be perfect. Formally, define $\pi^{\ast}$ as the vector of weights on seeds
that achieve perfect information aggregation, i.e., $\pi^{\ast}=\arg\min_{\pi
}var(\sum_{k}\pi_{k}x_{k})$, and let $v^{\ast}=var(\pi^{\ast}.x)$. We
have:\smallskip

\textbf{Result 1}: \textit{In the absence of transmission errors, the
equilibrium must be DG. In addition, in equilibrium, }$y_{i}=\pi^{\ast}%
.x$\textit{ and }$L_{i}=v^{\ast}$.\smallskip

In other words, as long as there is no noise, we get perfect agreement in
opinions in equilibrium and perfect information aggregation. As mentioned in
the introduction, the main difference with \cite{demarzo03} and \cite{golub10}
is that we allow for endogenous weights $\gamma_{i}$. For any connected
network, this is enough to obtain efficiency in equilibrium.

Intuitively, both $y_{i}$ and the neighbor's composite limit opinion
$\widehat{y}_{i}$ are weighted averages between $x_{i}$ and the composite seed
$\widehat{x}_{i}$, with different weights when players do not use DG rules. In
equilibrium, $i$ chooses optimally the weighting to reduce variance, so if the
equilibrium is not DG, the variance $v(y_{i})$ must be \textit{strictly}
smaller than the variance $v(\widehat{y}_{i})$, which itself is no larger than
the maximum variance $\max_{k}v(y_{k})$. Since this cannot be true for all
$i$, the equilibrium must be DG.


Regarding efficiency, in a DG equilibrium, player $i$ chooses the relative
weight $\pi_{i}$ on her own seed by modifying $\gamma_{i}$, and any departure
from perfect information aggregation leads $i$ to choose a relative weight
$\pi_{i}$ no smaller\textbf{ }than $\pi_{i}^{\ast}$. In a DG equilibrium,
$\pi_{i}$ also characterizes the influence of $x_{i}$ on the common long-run
opinion (there is consensus), so the weights $\pi_{i}$ must add up to 1. This
can only happen if they coincide with the efficient weights $\pi_{i}^{\ast}$.
Therefore there is a unique (and efficient) equilibrium outcome.

\subsection{Rule choice when there is noise}

We already saw that as soon as there is some noise, the outcome generated by
any DG rule drifts very far from minimizing $L_{i}.$ The loss grows without
bound. Indeed from the point of view of the individual decision maker it would
be better to ignore everyone else than to follow DG. In fact all strategies
that put too little weight on their own seed (recall DG puts zero weight) are
dominated from the point of view of the individual decision-maker, as well as
being socially suboptimal.\medskip{}

\textbf{Result 2:} \textit{Let }$\underline{m}=\varpi/(1+\varpi)$\textit{. Any
}$(m_{i},\gamma_{i})$\textit{ with }$m_{i}<\underline{m}$\textit{ is dominated
by }$(\underline{m},\gamma_{i})$\textit{, from the individual and social point
of view}.\medskip{}

Regarding the choice of the individually optimal rule, Result~2 builds on two
ideas. First, if all other players use DG, then for agent $i$, any $m_{i}>0$
is preferable to DG because everyone's opinion drifts off indefinitely if
$m_{i}=0$, as we saw above. Second, if some players use FJ (with $m_{j}>0$),
then initial opinions of these players $x_{j}$ (plus any persistent noise in
their reception of the signal) totally determines the long run outcome and the
seeds of all the players that use $DG$ do not get any weight --~they end up as
pure followers. This is not desirable for these DG players (and for the
others) for the same reason why, in the absence of noise, each one wishes to
let their own seed influence their long-run opinion. Hence the lower bound on
$m_{i}$.

To see why this is also true of the socially optimal rule, i.e. the rule that
minimizes $\sum_{i}L_{i},$ we observe that when $m_{i}=0$, the only effect of
information transmission by $i$ to his neighbors is to introduce $i$'s
perception errors into the network. When $i$ raises $m_{i}$ above $0$, he
raises the quality of the information he transmits, while limiting the
damaging echo effect that low $m_{i}$ generates. \textbf{\smallskip}

\subsection{How big is the divergence in opinions?\label{SectionDivergence}}

Result 2 has the obvious implication that full consensus is never going to be
an equilibrium when there are persistent errors--there are in fact two sources
of deviation, the error itself (which mechanically prevents consensus) and the
extra weight $m_{i}$ on one's initial signal (which fuels further
divergence.)\textbf{ }

Result 3 below shows that because of cumulated errors, the optimal weight put
on one's own seed tends to be relatively large, i.e. $O(\varpi^{1/3})$ (of the
order of $\varpi^{1/3}$).\footnote{When we say that $m=O(g(\varpi))$, we mean
that $m/g(\varpi)$ has a finite limit when $\varpi$ tends to $0$.} As a result
when $\varpi$ is small, the extra weight on one's own seeds becomes the
preponderant source of dispersion. These extra weights also determine the
equilibrium magnitude of $\widehat{\varpi}_{i}$ and $L_{i}$. We
have:\textbf{\smallskip}

\textbf{Result 3}: \textit{For any given finite network and any }$\varpi
>0$\textit{ small, all }$m_{i}$\textit{, }$p_{i}-\pi_{i}^{\ast}$,
$\widehat{\varpi}_{i}$ and $L_{i}-v_{i}^{\ast}$ \textit{are positive and
}$O(\varpi^{1/3})$ \textit{in equilibrium.}\smallskip

Note that in addition to cumulated errors, there is another source of
inefficiency in equilibrium, the fact that seeds are not efficiently weighted.
But that inefficiency is $O(\varpi^{2/3})$:\footnote{This is because for an
inefficient weighting of seeds $q\neq\pi^{\ast}$, the loss is second order in
the differences $q_{i}-\pi_{i}^{\ast}:$ $L_{i}-v^{\ast}=\sum(q_{i}^{2}-\pi
_{i}^{\ast2})\sigma_{i}^{2}=\sum(q_{i}-\pi_{i}^{\ast})^{2}\sigma_{i}^{2}%
+2\sum(q_{i}-\pi_{i}^{\ast})\pi_{i}^{\ast}\sigma_{i}^{2}$, and the last term
is $0$ because $\sum(q_{i}-\pi_{i}^{\ast})=1$ and at the optimum $\pi
_{i}^{\ast}\sigma_{i}^{2}=\pi_{j\text{.}}^{\ast}\sigma_{j}^{2}$ for all $i,j$}
a socially optimal choice of weights $m_{i}$ would trade-off more inefficient
weighting (larger $m$) against decreasing the variance of cumulated errors.


The intuition for Result 3 runs as follows. The error terms $\widehat{\varpi}$
are $O(\varpi/m^{2}).$ These errors terms degrade the quality of information
that each $i$ gets (raising $W_{i}$ above $W_{i}^{\ast}$), which in turn
implies a weighting $p_{i}$ of $i$'s seed larger than the efficient weighing
$\pi_{i}^{\ast}$, with $p_{i}-\pi_{i}^{\ast}$ at least $O(\varpi/m^{2})$ (by
(\ref{LP6}) and (\ref{LP6b})).
When $m>0$, players end up weighing seeds differently, but when all $m$ are
small, the spread between the weights is also small and $O(m)$. So if $p_{k}$
is the weight that $k$ puts on $x_{k}$, the weight that $i$ puts on $x_{k}$
must be $p_{k}+O(m)$. Since the weights that $i$ puts on all seeds must add to
1, the $p_{k}$'s must add up to at most $1+O(m)$. And since the sum $\sum
_{k}(p_{k}-\pi_{k}^{\ast})$ is at least $O(\varpi/m%
{{}^2}%
)$, $m$ must be at least $O(\varpi/m%
{{}^2}%
)$ in equilibrium, which gives $m$ at least $O(\varpi^{1/3})$.\footnote{The
proof also shows $m_{i}$ cannot increase beyond $O(\varpi^{1/3})$ in
equilibrium for the same reason that the equilibrium without error terms must
be DG: each player sets the weighting $p_{i}$ of own seed $x_{i}$ optimally,
and this creates a force towards optimal information aggregation.}

Note that Result~3 focuses on the case where variances are small.\textbf{
}When the $m_{i}$'s rise, the relative weights on seeds eventually diverge
sufficiently from efficient weighting that this fuels a further rise in
$W_{i}$ hence in $m_{i}$.

\subsection{Privately versus socially optimal choices}

We already showed that both private and social optima must deviate from DG
when there is noise. The next result shows that there is a sense in which, in
the presence of noise, the Nash Equilibrium is closer to DG than is desirable
from the point of view of social welfare maximization.\footnote{The result
shows that a marginal increase over equilibrium weights enhances welfare, but
we do not have a full characterization of socially efficient weights.}\medskip

\textbf{Result 4. }\textit{At any Nash equilibrium, a marginal increase of
}$m_{i}$ \textit{by any player }$i$\textit{ would increase aggregate social
welfare.}\medskip

To see why this result holds, assume $m_{j}\in(0,1)$ and observe that player
$j$'s opinion can be expressed as an average between the (modified) seeds
$\widetilde{x}_{-i}$ of players other than $i$ and player $i$'s opinion%
\begin{equation}
y_{j}=(1-\mu_{ji})C^{ji}\widetilde{x}_{-i}+\mu_{ji}y_{i} \label{eqinfluence}%
\end{equation}
where $C^{ji}$ is a probability vector and $\mu_{ji}\in(0,1)$,\footnote{This
assumes $m_{j}\in(0,1)$. $C_{j}^{ji}$ is positive because $j$ is using her own
seed.} with $\mu_{ji}$ and $C^{ji}$ both independent of $m_{i}$.

The expression above highlights that when player $i$ chooses $m_{i}$ optimally
(for him) to minimize the variance of $y_{i}$, there is no reason why he would
be also minimizing the variance of $y_{j}$. Specifically we use use
(\ref{eqinfluence}) to separate the loss $L_{j}$ into three terms:
\begin{equation}
L_{j}=(1-\mu_{ji})^{2}var(C^{ji}\widetilde{x}_{-i})+\mu_{ji}L_{i}+2(1-\mu
_{ji})\mu_{ji}Cov(C^{ji}\widetilde{x}_{-i},y_{i}). \label{eqCOV}%
\end{equation}
When $m_{i}$ is raised above $i^{\prime}s$ private optimum, there is no effect
on the first term. There is a second-order effect on the second term (because
we start at $i^{\prime}s$ private optimum). The last term is what creates a
discrepancy between private and social incentives.

This last term depends on the covariance between seeds other than that of $i$
($\widetilde{x}_{-i}$) and the opinion of $i$ ($y_{i}$).
When $m_{i}$ increases, the influence of each $k\neq i$ on $i$'s opinion is
reduced, and the correlation between $y_{i}$ and $x_{k}$ (and even more so
with $\widetilde{x}_{k}$) is also reduced. Therefore, starting at a Nash
equilibrium, $L_{j}$ goes down when $m_{i}$ is raised.

\subsection{Simple examples.}

To conclude this Section, we directly compute the equilibrium and socially
efficient weights in simple examples to shed further light on the rule choice
and information aggregation. We assume that initial opinions are equally
informative ($\sigma_{i}^{2}=1$ for all $i$) and each player treats all his
neighbors symmetrically ($A_{ij}=1/|N_{i}|$). We start with the two-player
network and next discuss other larger simple networks (directed circle,
complete network and star network).

\subsubsection{The two-player case}

\paragraph{Social optimum.}

Assuming independent errors, we obtain from (\ref{eq2})
\[
L_{1}=I(p_{1})+(p_{1})^{2}\mathcal{X}(m_{1})+(1-p_{1})^{2}\mathcal{X}(m_{2})
\]
where $I(p)=p^{2}+(1-p)^{2}$ is the variance of long run opinion in the
absence of transmission noise and $\mathcal{X}(m)=\varpi\frac{(1-m)^{2}}%
{m^{2}}$ represents the effect of cumulated noise. The total social loss is
$L=L_{1}+L_{2}.$

It is easy to check that, given the symmetry, minimizing the social loss
requires setting identical values for $m_{1}$ and $m_{2}$. When both players
use the same rule ($m=m_{1}=m_{2})$, $p_{i}=\frac{1}{2-m}$ and the social loss
is:
\[
L=2I(\frac{1}{2-m})(1+\mathcal{X}(m))
\]
The expression highlights a trade-off between decreasing $m$ for information
aggregation purposes ($I(p)$ is minimized at $p=1/2$), and increasing $m$ to
limit the effect of cumulated communication errors (when $\varpi>0$ and $m$ is
small, communication errors are hugely amplified).


Welfare is maximized for an $m^{\ast\ast}$ that optimally trades off these two
effects and the socially efficient weight $m^{\ast\ast}$ (which minimizes $L$)
can be significantly different from $0$ even when $\varpi$ is small.
Specifically, for $\varpi=0.0001$, $m^{\ast\ast}=0.13$ and for $\varpi=0.001$,
$m^{\ast\ast}=0.21$. Furthermore, for $\varpi$ small, $m^{\ast\ast}%
\simeq(4\varpi)^{1/4}$.\footnote{This is because for $\varpi$ small
$L\simeq1+$ $m^{2}/4+\varpi/m^{2}$.}

\paragraph{Nash Equilibrium.}

We now assume that individuals choose their rules non-cooperatively. Applying
Proposition~5, we obtain $p_{i}/(1-p_{i})=1+\widehat{\varpi}_{i},$ so
\[
p_{i}=\frac{1+\widehat{\varpi}_{i}}{2+\widehat{\varpi}_{i}}\text{ where
}\widehat{\varpi}_{i}=E\widehat{\xi}_{i}^{2}%
\]
which gives the best response for $i$, as a function of $m_{j}$:%
\[
m_{i}=\frac{m_{j}(1+\widehat{\varpi}_{i})}{1+m_{j}(1+\widehat{\varpi}_{i})}%
\]
Figure 1 plots the best responses for $\varpi=0.01$.

\begin{figure}[h]
\centering
\includegraphics[scale=0.4]{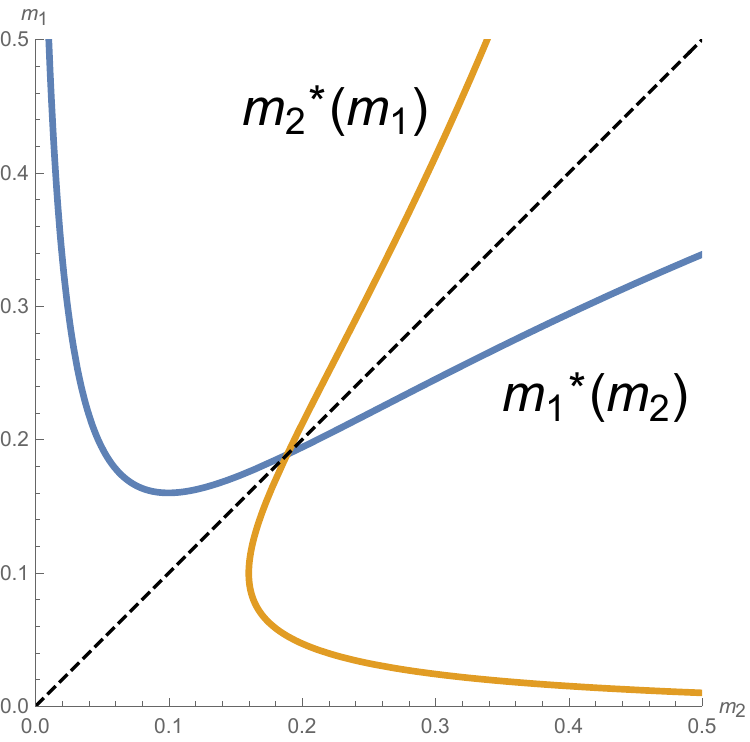}\caption{Best responses,
$\varpi=0.01$}%
\end{figure}In the absence of noise, $\widehat{\varpi}_{i}=0$, and player 1
should set $m_{1}$ so that $p_{1}=1/2$ (for information aggregation purposes),
which requires $m_{1}<m_{2}$, which explains why there is no equilibrium with
positive $m$ (this is the force towards DG). With noise, the variance
$\widehat{\varpi}_{i}$ explodes when $m_{j}$ gets small, reflecting the
cumulation of errors when $m_{j}$ is low. This provides $i$ with incentives to
raise $p_{i}$ (hence $m_{i}$) which in turn puts a lower bound equilibrium
weights: in equilibrium, $m_{1}^{\ast}=m_{2}^{\ast}=m^{\ast}$ and $m^{\ast}$
is a solution to
\[
m^{\ast}=\frac{\widehat{\varpi}^{\ast}}{1+\widehat{\varpi}^{\ast}}\text{ with
}\ \widehat{\varpi}^{\ast}=\varpi\frac{1+(1-m^{\ast})^{2}}{m^{\ast2}}.
\]
When $\varpi$ is small, we have $m^{\ast}\simeq(2\varpi)^{1/3}$. Since
$m^{\ast\ast}\simeq(4\varpi)^{1/4}$, the ratio of $m^{\ast\ast}$ to $m^{\ast}$
become arbitrarily large when $\varpi$ is small.

\subsubsection{Larger networks.}

Equilibrium weights are obtained using Proposition~6: player $i$'s incentive
condition yields
\[
\frac{m_{i}}{1-m_{i}}=W_{i}/h_{i}%
\]
where $W_{i}=var(\widehat{x}_{i})+\widehat{\varpi}_{i}$. Both $h_{i}$ and
$W_{i}$ depend only $m_{-i}$ and the structure of the network, and the
equilibrium values $m_{i}^{\ast}$ are obtained by simultaneously solving these
equations. For the directed circle and complete network, all players are
symmetric so one easily finds the equilibrium weight $m^{\ast}$. For the star
network, we have to examine the incentives of the central player (labelled
player $0$) and peripherical players separately, and we obtain equilibrium
weights $m_{0}^{\ast}$ and $m^{\ast}$ for central and peripherical players
respectively. We leave the details of the computation to the Appendix,
focusing on the case where the variance $\varpi$ is small. We report here some
notable facts.

In the star network, the central player can have disproportionate influence on
the opinions of others, and in equilibrium, she chooses $m_{0}^{\ast}$ much
below $m^{\ast}$ to ensure that this is not the case: for fixed $n$,
$\frac{m_{0}^{\ast}}{m^{\ast}}\simeq\frac{1}{n-1}$, and at the large $n$
limit, $\frac{m_{0}^{\ast}}{m^{\ast}}\simeq O(\varpi^{1/3})$. So in effect, at
this limit, the central player essentially ignores her signal and behaves like
a DG player.

It is also interesting to compare the aggregation properties of different
networks. Both $h_{i}$ and $W_{i}$ depend on the structure of the network, and
this eventually affects the performance of the network. For example, for fixed
$m$, the cumulated error terms $\widehat{\varpi}_{i}$ are higher in the star
network than in both other networks, because the central player's error term
contaminates all other players in a correlated way. The consequence is that
the star network performs worse than both the directed circle and the complete network.

Finally, we find that the directed circle performs better than the complete
network when $n$ is not too large, because, while the terms $W_{i}$ are
similar across these two networks, players have stronger incentives to raise
$m_{i}$ in the directed network.\footnote{Technically, this is because $h_{i}$
is smaller in the directed network, so a given target $p_{i}$ is achieved with
a higher $m_{i}$.} The comparison is reversed for large $n$--the directed
circle yields poorer information aggregation ($var(\widehat{x}_{i})$ is
higher) and, when persistent errors are independent, poorer averaging of errors.


\section{\textbf{Implications for the divergence of opinions and
polarization\label{SectionSimple}}}

In the absence of noise, and if players use DG with appropriate weights
$\gamma$, long-run opinions converge to a consensus $y^{\ast}=\pi^{\ast}.x$
which efficiently aggregates seeds. In a large network, this opinion $y^{\ast
}$ will essentially coincide with the underlying state $\theta$ ($y^{\ast
}\simeq\theta$), which implies that if we consider two such identical
networks, there will be \textit{consensus within} each network and
\textit{consensus across} networks.

In the presence of noise, two things may happen. A divergence of long-run
opinions $y$ away from $y^{\ast}$, which means a \textit{divergence of average
opinions between the networks}, as well as some \textit{dispersion of opinions
within networks}. This section argues that there is a connection between
consensus within subgroups (low dispersion) and polarization (high divergence
across subgroups).


To fix ideas, we consider below the case of two large disconnected star
networks modeled as above.\footnote{Result 5 below would also hold if the set
of cross-star links were a vanishingly small proportion of the total number of
links.} This description generally fits the maps of social networks in the US
population with the two stars representing Democrats and Republicans
(\cite{cox20}). We assume that in each star network all peripheral players use
the same weight $m$ and that central players behave as DG players, just
aggregating peripheral players' opinions.\footnote{In the Appendix, we
consider the case where central players benevolently choose $m_{0}$ to
minimize the losses of the peripheral players, given $m$.} We are interested
in the effect on $m$ on the distribution of opinions within the star and
across stars. In each star, if the central player is labelled player~$0$, for
any peripheral player $i$ of that star we have:%
\[
y_{i}=mx_{i}+(1-m)(y_{0}+\xi_{i})
\]
The dispersion of opinions between two peripheral players \textit{within} a
given star is
\[
d\equiv E(y_{i}-y_{j})^{2}=2(m)^{2}+2(1-m)^{2}\varpi
\]
The \textit{average} opinion of peripherical players is $\overline
{y}=m\overline{x}+(1-m)(y_{0}+\overline{\xi})$, and for a large network, with
independent errors, only $y_{0}$ contributes to the variance of $\overline{y}%
$. \textit{Across} the networks, average opinions are independent (conditional
on $\theta$) and the dispersion of opinion $D$ between average opinions is
thus:
\[
D=2v(\overline{y})
\]
The following result establishes a relationship between $d$ and $D$, where
$\varpi_{0}$ refers to the variance of $\xi_{0}$:\medskip

\textbf{Result 5:} \textit{Fix }$\varpi_{0}$\textit{ small and assume
independent errors. At the social optimum }$m^{\ast\ast}$\textit{, }$D\simeq
d$\textit{ and for any }$m\leq m^{\ast\ast}$\textit{, }$D\simeq\frac
{4\varpi_{0}}{d}$\textit{.}\medskip

\textbf{Proof}: When the central player is DG, $y_{0}=\overline{y}+\xi_{0}$,
so for a large network and independent errors this immediately gives
$\overline{y}=\frac{(1-m)\xi_{0}}{m},$ hence $D\simeq\frac{2\varpi_{0}}{m^{2}%
}\simeq\frac{4\varpi_{0}}{d}$ for small $m$. Writing $y_{i}=(y_{i}%
-\overline{y})+\overline{y}$, we obtain $v(y_{i})=\frac{1}{2}(d+D)$. Since
$D\simeq\frac{4\varpi_{0}}{d}$, the loss $v(y_{i})$ is minimized for $D\simeq
d\simeq2\varpi_{0}^{1/2}$ (hence $m^{\ast\ast}\simeq\varpi_{0}^{1/4}%
$).$\blacksquare$

Result~5 says that the social optimum is achieved for $D\simeq d$ and it
establishes a relationship between consensus within each group (small $d$) and
polarization across groups (high $D$): as $m$ decreases below $m^{\ast\ast}$,
within-group consensus goes up but so does polarization across the groups.

Our equilibrium analysis provides one possible reason for $m$ being too low,
but there may be others. For example, imagine that for some issues, the errors
$\xi_{i}$ are correlated across network members (calling for higher $m$),
while for other issues, the errors are independent (calling for lower $m$). If
agents are unable to adjust $m$ to the type of problem they face, the weights
$m$ will be inefficiently low for the problems where there are correlated
errors, thus fostering too much consensus and polarization for these
problems.

\section{Idiosyncratic errors}

We now introduce idiosyncratic errors and assume that
\[
\varepsilon_{i}^{t}=\xi_{i}+\nu_{i}^{t}%
\]
where $\nu_{i}^{t}$ are i.i.d. across individuals and
time.\footnote{Implicitly, we think of $\nu_{i}^{t}$ as an error in
interpreting the opinions expressed by others. Alternatively, one could
consider errors in expressing one's opinion.} We further assume $E\nu_{i}%
^{t}=0$ and let $\varpi^{0}=var(\nu_{i}^{t})$. We wish to characterize the
(additional) loss generated by these idiosyncratic errors, and examine the
consequence regarding incentives.

In the absence of idiosyncratic elements, the speeds of adjustment $\gamma
_{i}$ plays no role when $m_{i_{0}}>0$ for $i_{0}$. The main insight of this
Section is that idiosyncratic errors induce temporary variations in opinions
which are potentially costly, and players have incentives to reduce these
variations by decreasing $\gamma_{i}$. Furthermore, when all players choose an
arbitrarily small $\gamma_{i}$, long-run opinions essentially coincide with
the ones obtained in the absence of idiosyncratic errors.

Formally, for any fixed $m$, $x$ and $\xi$, we define the expected opinion
vector $\overline{y}^{t}=Ey^{t}$ where the expectation is taken over all
$\nu_{i}^{s}$ for $s\leq t$. We also let $\eta_{t}=y^{t}-\overline{y}^{t}$ and
$V^{t}=var(\eta_{t})$. Furthermore, we let $y^{0}$ denote the long-run opinion
that would obtain \textit{in the absence of idiosyncratic errors}, and
$L_{i}^{0}=var(y^{0})$ the associated loss of player $i$ computed over
realizations of $x$ and $\xi$. The next Proposition (proved in Appendix~B)
provides the analog of Propositions 1 to 3 for the idiosyncratic noise
case:\smallskip

\textbf{Proposition 7}: \textit{If }$m_{i}=0$\textit{ for all }$i$\textit{,
}$V^{t}$\textit{ increases without bound.} If\textit{ }$m_{i_{0}}>0$\textit{
for some }$i_{0}$,\textit{ }$\overline{y}^{t}$\textit{ and }$V^{t}$\textit{
both have well-defined limits }$\overline{y}$\textit{ and }$V$\textit{.
Besides, }$\overline{y}=y^{0}$\textit{, }$V$\textit{ is independent of }%
$x$\textit{ and }$\xi$\textit{, and }$L_{i}=L_{i}^{0}+V$. \textit{Furthermore,
if }$m_{i}\leq\overline{m}$ and $\gamma_{i}\geq\underline{\gamma}$ for all
$i$, $V_{i}\geq\frac{\varpi^{0}}{2n}\frac{\underline{\gamma}^{2}%
(1-\overline{m})^{2}}{\overline{m}}.$\smallskip

For given $m$, $\gamma>0$, expected long-run opinions eventually coincide with
$y_{0}$, but long-run opinions are subject to temporary changes resulting from
idiosyncratic communication errors. Proposition~7 shows that, for given
$\gamma$, these temporary changes are significant and costly and when all $m$
are small.\smallskip

However choosing a lower $\gamma_{i}$ slows down the adjustment of one's
opinion. Result~6 below shows that for small enough $\gamma_{i}$, long-run
opinions becomes essentially unaffected by temporary shocks in perceptions or
temporary variations in others' opinions. \smallskip

\textbf{Result 6:} \textit{Fix} $\underline{m}$. \textit{We have:}
\newline\textit{(i)There exists }$c$ \textit{such that for any} $\gamma>0$ and
$m\geq\underline{m}$, $V_{i}\leq c\max\gamma_{j}$. \newline\textit{(ii) For
any }$\gamma_{-i}>0$\textit{, there exists }$c$\textit{ such that for all
}$m\geq\underline{m}$, $V_{i}\leq c\gamma_{i}$. \smallskip


The proof is in Appendix B. Item (i) shows that when all $\gamma_{i}$ are
small, all $V_{i}$ are small. Item (ii) shows that by choosing $\gamma_{i}$
very small, a player can get rid of the additional variance induced by the
idiosyncratic noise.


Note that the incentive to set $\gamma_{i}$ arbitrarily small obviously
depends on the assumption that players only care about long-run opinions. If
players also cared about opinions at shorter horizons, then they would have
incentives to increase $\gamma_{i}$ to more quickly absorb information from
the opinions of others: the trade-off is between increasing the rate of
convergence (which is desirable when the relevant horizon is shorter) and
increasing the variance induced by idiosyncratic noise (which is not desirable).

\section{Discussion}

We discuss various extensions of and possible variations upon our base model.
We examine the case of biases persistent errors, showing that this provides
additional incentives to raise $m_{i}$ and increases the losses $L_{i}$. We
show that FJ rules are robust to variations in the communication protocol. We
also discuss how non-stationary rules might create further difficulties. We
also examine the case of coarse communication, which enables us to discuss the
relationship between our work and recent papers on information aggregation in
network when agents' priors are misspecified (\cite{frick20}, \cite{bohren21}%
). The connection with that literature is that our persistent errors play the
same role as a misspecification. One key difference is that we allow players
to correct, to some extent (i.e., through the weight $m_{i}$), for the
mispecifications that players are subject to.

\subsection{Biased persistent errors}


We have so far assumed that the persistent error is drawn from a distribution
that is mean zero. One can however imagine settings where it is reasonable to
assume that the persistent error is biased, centered on $\xi_{i}^{0}$ for
player $i$. This could be because some individuals are systematically biased
in what they report or process (for whatever reason), or because others
erroneously believe that they are and wrongly correct for it. Another reason
could be that preferences are heterogenous, say each person cares about
$\theta_{i}=\theta+b_{i}$, observes $x_{i}=\theta_{i}+\delta_{i}$, but has an
imprecise and potentially biased estimate of the vector of preference spreads
$\beta_{i}=(b_{j}-b_{i})_{j}$.

In either case, adding systematic biases $\xi_{i}^{0}$ can only raise the
terms $\widehat{\varpi}_{i}=E\widehat{\xi}_{i}$, thus providing additional
incentives to raise $m_{i}$ and increasing the losses $L_{i}$.

\subsection{Other communication protocols\label{SectionProtocols}{}}

We have followed the standard approach to modeling communication in this
literature, with each player communicating with all his neighbors at every
date.\footnote{\cite{banerjee21} introduce the idea of a Generalized
DeGroot model where not everyone starts with a signal and therefore does not
participate in the communication till they get a signal. They show that this
partially weakens the "wisdom of crowds".} We now consider an extension where
each round of communication is one-sided and, at any date $t$, each agent $i$
only hears from a subset $N_{i}^{t}\subset N_{i}$ of his neighbors but there
exists $K$ such that each player hears from all his neighbors at least once
every $K$ periods.\footnote{\label{footnoteA}That is, for all $t:\cup
_{s=1,..,K}N_{i}^{t+s-1}=N_{i}.$} Imperfect communication is modeled as
before, through the addition of an error term $\xi_{i}$ that slants what $i$
hears. Together these give us%
\begin{align*}
z_{i,j}^{t}  &  =y_{j}^{t-1}+\xi_{i}\text{ if }j\in N_{i}^{t}\\
z_{i,j}^{t}  &  =z_{i,j}^{t-1}\text{ if }j\in N_{i}\backslash N_{i}^{t}%
\end{align*}
where $z_{i,j}^{t}$ is $i$'s current perception of $j$'s opinion, based on the
last time he has heard from $j$. Player $i$ uses these perceptions to
construct an average over neighbor's opinions
\[
z_{i}^{t}=A_{i}Z_{i}^{t}%
\]
where $Z_{i}^{t}=(z_{i,j}^{t})_{j}$ is the vector of $i$'s perceptions and
$A_{i}=(A_{ij})_{j}$ defines as before how $i$ averages others' opinions. We
continue to assume FJ updating. We have:\smallskip\smallskip

\textbf{Proposition 8}: \textit{Assume at least one player, say }$i_{0}%
$\textit{, updates according to FJ with }$m_{i_{0}}>0$\textit{. Then for any
fixed }$x,\xi$\textit{, }$y^{t}$\textit{ converges and the limit vector of
expected opinions }$y$\textit{ is independent of the protocol.}\footnote{So
long as the condition in footnote \ref{footnoteA} holds.}\smallskip\smallskip

Intuitively, convergence obtains for standard reasons, and at the limit, since
expected opinions do not change, the timing with which one hears others does
not matter (see Appendix).

This robustness contrasts with what happens when players use DG rules.
For example, consider two agents using DG rules and assume that agent 1
updates every period, while agent 2 updates every three other periods. At
dates $t$ where 2 updates, we have:
\begin{align*}
y_{1}^{t}  &  =(1-\gamma_{1})^{3}y_{1}^{t-3}+(1-(1-\gamma_{1})^{3})y_{2}%
^{t-3}\\
y_{2}^{t}  &  =(1-\gamma_{2})y_{2}^{t-3}+\gamma_{2}y_{1}^{t-1}\\
&  =(1-\gamma_{2})y_{2}^{t-3}+\gamma_{2}((1-\gamma_{1})^{2}y_{1}%
^{t-3}+(1-(1-\gamma_{1})^{2})y_{2}^{t-3})\\
&  =(1-\gamma_{2}(1-\gamma_{1})^{2})y_{2}^{t-3}+\gamma_{2}(1-\gamma_{1}%
)^{2}y_{1}^{t-3}%
\end{align*}
So, the process evolves as if weights were $\gamma_{1}^{\prime}=1-(1-\gamma
_{1})^{3}>\gamma_{1}$ and $\gamma_{2}^{\prime}=\gamma_{2}(1-\gamma_{1}%
)^{2}<\gamma_{2}$. This means that with DG rules, changes in the frequencies
with which players communicate amount to changes in the values of $\gamma_{i}$
(when you hear less often from others, your opinion changes more slowly,
effectively reducing $\gamma_{i}$). And even when communication is noiseless,
these changes modify long-run opinions: if $\gamma_{i}$ goes down, long-run
opinions get closer to $i$'s opinions (see Section~\ref{sectionDGwithout}
Equation~(\ref{pi})).

Thus, even in the absence of transmission errors, variations in the
communication protocol induce additional variation in long-run opinions which
can be mitigated by the use of FJ rules by all players. That said, in the
absence of transmission errors, long-run opinions under DG remain averages
over initial opinions, so the fragility is not as severe as the one already
highlighted: the variance induced by variations in the protocol remains
bounded even when $m_{i}=0$.

\subsection{Coarse communication\label{SectionCoarse}}

In the social learning literature, it is common to focus on choice problems
where there are two possible actions, and the information being aggregated is
which of the two is being recommended by others. Coarse communication is
potentially a source of herding, but when agents have many neighbors, the
fraction of players choosing a given action may become an accurate signal of
the underlying state. We explain below how our model can accommodate an
economic environment of this kind, and we use this to relate our findings to
\cite{ellison93,ellison95} and \cite{frick20}, as well as \cite{bohren21}.

Assume heterogenous preferences with $\theta_{i}=\theta+b_{i}$ characterizing
$i$'s value from choosing $1$ over $0$, so the optimal action $a_{i}^{\ast}$
is $1$ when $\theta_{i}>0$, $0$ otherwise.\footnote{Thus for $i$ with
preference parameter $b_{i}$,\ choosing $0$ when $\theta+b_{i}>0$ costs
$\theta+b_{i}$. When agents choose between products 1 or 0, $\theta$
represents a relative quality dimension affecting all preferences, as in
\cite{ellison93}.} Agent $i$ knows $b_{i}$ but does not know $\theta$
perfectly. He has an initial opinion $x_{i}=\theta+\delta_{i}$ and aggregates
opinions of others to sharpen his assessment of $\theta$. Assume the $b_{i}$'s
are drawn from identical distribution $g$ (and cumulative denoted $G$) with
full support on $\mathcal{R}$.

We define, as before, $y_{i}^{t}$ as agent $i$'s opinion (about $\theta$) at
date $t$ and we assume that an agent with current opinion $y_{i}^{t}$ reports
$a_{i}^{t}=1$ if $y_{i}^{t}+b_{i}>0$ and $a_{i}^{t}=0$ otherwise. Each agent
$i$ observes the fraction $f_{i}^{t}$ of neighbors that choose action $0$,
which she can use to make an inference $\psi_{i}(f_{i}^{t})$ about $\theta$,
and update her opinion using an FJ-like rule:
\[
y_{i}^{t+1}=(1-\gamma_{i})y_{i}^{t}+\gamma_{i}(m_{i}x_{i}^{t}+(1-m_{i}%
)\psi_{i}(f_{i}^{t}))
\]

Long-run opinions clearly depend on the inference rule assumed, but there is a
natural candidate for $\psi_{i}$, the function $\phi\equiv h^{-1}$, where
$h(y)\equiv G(-y)=\Pr(y+b_{i}<0)$ is the fraction of agents that choose $a=0$
when their opinions are all equal to $y$. If others have opinions that are
correct and equal to $\theta$, a fraction $f\simeq h(\theta)$ choose $a=0$ and
$h^{-1}(f)$ is a good proxy for $\theta$. Of course this assumes that agents
know the distribution over preferences. In the spirit of our previous
analysis, let's assume that%
\[
\psi_{i}(f)=\phi(f)+\xi_{i}%
\]
where $\xi_{i}$ is a persistent error in interpreting $f$.\footnote{As in
\cite{frick20}, $\xi_{i}$ could arise from an erroneous prior $g_{i}\neq g$,
with agents using the inference function $\psi_{i}=h_{i}^{-1}$ where
$h_{i}(\theta)=G_{i}(-\theta)$. The difference $\xi_{i}(f)\equiv\psi
(f)-\phi(f)$ is an error in making inferences. With preferences centered on
$\overline{b}$, and agent having an erroneously translated prior centered on
$\overline{b}_{i}$, the error is independent of $f$ and equal to $\xi
_{i}\equiv\overline{b}_{i}-\overline{b}$.}$^{,}$%
\footnote{\citet[Section~1]{ellison93} examines social learning assuming
$b_{i}=0$ for all and $\psi_{i}(f)=f-1/2$: choices are tilted in favor of the
more popular one. EF find that small enough $m$s generate perfect learning in
the long-run. A key aspect of the inference rule $\psi_{i}(f)$ is that it
correctly maps the sign of $f-1/2$ to the sign of $\theta$, which, given
homogeneity, is the only thing that agents care about. (Note that in EF,
agents receive many signals $x_{i}$ about the state, but, given their
assumptions, their model is equivalent to the one proposed here where agents
just receive one signal at the start).} To fix ideas, we assume correlated
errors ($\xi_{i}=\xi$ for all $i$) with variance $\varpi$.

Within this extension, we may ask about the fragility of long-run opinions
when $m$ is small, as well as equilibrium and socially efficient weights
(details are provided in the Appendix).

DG-like rules ($m=0$) generate long-run opinions unanimously in favor of $a=1$
if $\xi>0$, $a=0$ if $\xi<0$, \textit{independently of the underlying state
and the initial signals received.}

Under FJ with small $m$, long-run opinions remain anchored on initial
opinions, but long run opinions drift away from $\theta$ and converge to
$\theta+\frac{(1-m)\xi}{m}$. The trade-off is thus similar to the one in our
basic model. Raising $m$ reduces fragility with respect to transmission noise,
dampening the echo term $\frac{(1-m)\xi}{m}$. And agents continue to diagree
even in the long-run. The consequence regarding social incentives and private
incentives is as before, with $m^{\ast}$ and $m^{\ast\ast}$ respectively
comparable to $\varpi^{1/3}$ and $\varpi^{1/4}$: agents do not incorporate the
damaging echo effect that an $m_{i}$ set too low produces in their choice of
$m_{i}$ .


\subsection{A connection to misspecified Bayesian
models.\label{subsectionMisspecified}}

How do the results in the previous sub-section relate to the results from
Bayesian models where agents have misspecified priors (and in particular
\cite{frick20} and \cite{bohren21})? Consider a social learning environment
related to these Bayesian models where players move in sequence and observe
all previous choices. Preferences and signals are as defined above. Assume the
true state is $\theta_{0}$. Under Bayesian learning, if beliefs get highly
concentrated on some $\theta$, then private signals do not affect decisions
much and the fraction $f$ of people that choose $a=0$ are approximately those
for which $\theta+b<0$ so $f\simeq G(-\theta)$. If agents have an erroneous
prior about the distribution of $b$'s and believe its cumulative is shifted by
$\xi$ (say, $\widehat{G}(b)\equiv G(b-\xi)$) then agents are expecting a
fraction close to $\widehat{f}=\widehat{G}(-\theta)=G(-\theta-\xi)$, so if
$\xi>0,$ $\widehat{f}<f$. When the subjective prior over states has full
support, this should inevitably lead agents to believe that the state is lower
than $\theta$ (to justify the higher-than-expected $f$ observed) and so on...,
which explains the fragility result obtained in \cite{frick20}).

Let us now introduce, as in \cite{bohren21}, a fraction $q$ of autarkic
players that only base their choice on their private signal $x_{i}$ (thus
ignoring the social information). Define $G^{0}(\theta)$ as the fraction of
autarkic types that choose $a=0$ when the state is $\theta$, and to fix ideas,
further assume that non-autarkic types have correct priors about $G^{0}$. When
beliefs of non-autarkic types are concentrated on $\theta$ and the true state
is $\theta_{0}$, the fraction $f$ becomes
\[
f=qG^{0}(-\theta_{0})+(1-q)G(-\theta)
\]
while a fraction
\[
\widehat{f}=qG^{0}(-\theta)+(1-q)G(-\theta-\xi)
\]
would be expected. The observed $f$ will meet expectations when
\[
G^{0}(-\theta)-G^{0}(-\theta_{0})=\frac{1-q}{q}(G(-\theta)-G(-\theta-\xi))
\]
which implies a discrepancy $\Delta=\theta_{0}-\theta$ comparable to
$\frac{\xi}{q}$, which thus blows up when $q$ is small.

To relate this to our paper, observe that a measure $q$ of autarkic types
generates an overall inefficiency comparable to $q$ (because they are not
using information so each experiences a loss comparable to $1$), while when
$\xi$ is a random variable with variance $\varpi$, the loss induced by the
discrepancy $\Delta$ is quadratic in $\Delta$, so comparable to $\frac{\varpi
}{q^{2}}$, which in turn implies that to implement the social optimum (to
minimize the overall loss), $q$ should be comparable to $\varpi^{1/3}$.

Autarkic types thus play a role similar to our weights $m_{i}$, helping the
anchoring the beliefs of social types.\footnote{Note that unlike Bohren and
Hauser, we find here that the fraction $q$ needs to be large enough. This is
because, unlike BH who assume few states and correct priors over states, we
assumed here that subjective priors on $\theta$ have full support.} In our
setup, the analog of social and autarkic types would be to assume that agents
are either $DG$ ($m_{i}=0$) or use $m_{i}=1$. In contrast, we have assumed
that some intermediate $m_{i}$ is feasible for each agent.

Another difference is that we focus on the optimal choices of $m_{i}$ from the
social or private points of view. In looking for a Nash equilibrium, we
decentralize the choice of $m_{i}$ and endogenize the weight each puts on
social versus private information.\footnote{A similar decentralization
exercise (endogenizing $q$) could be done in the BH environment with agents
choosing ex ante whether to be autarkic or social, with the consequence that
in equilibrium, they would have to be indifferent between the two roles, hence
incur a significant loss (equal to that of the autarkic type).}

The lesson we draw from this discussion is that both DG and Bayesian updating
are sensitive to transmission or specification errors for a similar reason:
they both incorporate a force towards consensus, but since consensus is not
feasible (because of the errors), beliefs end up being pushed to the
boundaries of the feasible set of states. FJ-like rules, to the extent that
they allow for sufficiently diverse opinions or beliefs, end up being more
robust.

\subsection{Uncertainty over the precision of initial
signals.\label{SectionUncertainty}{}}

We examine here another variation of the model, assuming that the precision of
initial signals is a random variable and that players are able to ajust the
speed $\gamma_{i}$ as a (linear) function of $\sigma_{i}^{2}$. We argue below
that in the absence of processing errors, this type of shock does not affect
the performance of DG and therefore, unlike where there are errors, there is
no incentive for players to use the instrument $m_{i}$.

Formally, assume that each the speed of adjustment $\gamma_{i}$ as a linear
function of the variance of signal, that is, $\gamma_{i}=\mu_{i}\sigma_{i}%
^{2}$. Then for well-suited coefficients $\mu^{\ast}=(\mu_{i}^{\ast})_{i}$
information aggregation is perfect, which further implies that this particular
$\mu^{\ast}$ is also a Nash Equilibrium of the game where each chooses
$\mu_{i}$.

To see why, recall that under DG, the consensual long-run opinion is a
weighted average of initial opinions, with weights proportional to $\rho
_{i}/\gamma_{i}$ (see (\ref{pi})). So if the $\mu_{i}$'s are proportional to
$\rho_{i}$, the weights become proportional to $\rho_{i}/\gamma_{i}$, hence
proportional to $1/\sigma_{i}^{2}$, implying that perfect aggregation obtains
for each vector of realization $(\sigma_{1},...,\sigma_{n})$.

\subsection{Non-stationary weights.\label{SectionNonStationary}}

The updating processes that we consider have stationary weights. Agents do not
attempt to exploit the possibility that early reports possibly reveal more
information than later reports: later reports from neighbors may incorporate
information that one has oneself transmitted to the network, and therefore
should have lesser impact on own opinion.

As a matter of fact, with two players, one could imagine a process in which
(i) player 1 combines the first report he gets with own opinion, yielding
$y_{1}=m_{1}x_{1}+(1-m_{1})(x_{2}+\varepsilon)$, and then ignores any further
reports from player 2; and (ii) player 2 follows DG. With $m_{1}$ set
appropriately, such a process would permit player 1 to almost perfectly
aggregate information and player 2 to benefit from that information
aggregation performed by player 1.

There are however important issues with such time-dependent processes. In
particular, it is not obvious how one extends these to larger networks since
they require that each person knows his or her role in the network. They are
also sensitive to the timing with which information gets transmitted or heard.
With some randomness in the process of transmission, it could for example be
that the first report $y_{2}$ that player 1 hears already incorporates player
1's own signal (because after a while $y_{2}$ starts being a mixture between
$x_{2}$ and $x_{1}$), and as a result, player 1 should put more weight on the
opinions of others. But of course, in events where $y_{2}=x_{2}$, this
increase in weight makes information aggregation worse.

To illustrate this strategic difficulty in a simple model with noisy
transmission, assume that time is continuous, communication is\textbf{\ }%
one-sided (either 1-%
$>$%
2 or 2-%
$>$%
1), with each player getting opportunities to communicate at random dates. The
processes generating such opportunities are assumed to be two independent
Poisson process with (identical) parameter $\lambda$. Also assume that a
report, once sent, gets to the other with probability $p$. Consider the
time-dependent rule where each person communicates own current opinion, and
their current opinion coincides with their initial opinion if one has not
received any report ($y_{i}=x_{i}$), and otherwise coincides with $y_{i}%
=m_{i}x_{i}+(1-m_{i})z_{i}^{f}$ where $z_{i}^{f}$ is the perception of the
first report received. Even if perceptions are almost correct (i.e.
perceptions almost coincide with the other's current opinion), the noise
induced by the communication channel generates uncertainty about who updates
first, contributing to variance in the final opinion for all $m_{i}$. For
example, in events where player 1 already sent a report and receives one from
player 2, it matters whether player 2 received the report that 1 sent and
incorporated it into her opinion, or whether player 2 failed to receive the
report, in which case what player 1 gets is player 2's initial opinion.

In contrast, the time-independent FJ is not sensitive to that noise and
achieves reasonably good information aggregation for many values of
$m=m_{1}=m_{2}$. FJ rules conveniently address a key issue in networks:
whether what I hear already incorporates some of what I said.{}

\section{Concluding remarks}

We end the paper with a discussion of issues that we have not dealt with, and
which may provide fruitful directions for future research.

One premise of our model is that everyone has a well-defined initial signal.
However the analysis here would be essentially unchanged if some players did
not have an initial opinion to feed the network and were thus setting
$m_{i}=0$ for the entire process. FJ would aggregate the initial opinions of
those who have one.

In real life many of our opinions come from others and in ways that we are not
necessarily aware of, and the existence of a well-defined ``initial opinion"
could be legitimately challenged. In other words, people may have a choice
over the particular opinion they want to hold on to and refer back to (in
other words, the one that gets the weight $m_{i})$.

To see why this might matter, consider a variation of our model where some
players ($N^{dg}$) have initial opinions but use DG rule (or set $m_{i}$ very
low), while other agents $(N^{fj})$ have no initial opinions (or very
unreliable ones). In this environment, there is a risk that the initial
opinions of the $DG$ players eventually disappear from the system, and soon
are overwhelmed by noise in transmission. The other (non-DG) players could
provide the system with the necessary memory, using the initial communication
phase to gradually build up an ``initial opinion" based on the reports of
their more knowledgeable DG neighbors, and then seed in perpetually that
``initial opinion" into the network. In other words, in an environment where
information is heterogeneous and weights $m_{i}$ are set sub-optimally by
some, there could be a value for some agent in adopting a more sophisticated
strategy in which the ``initial opinion" is updated for some period of time
before it becomes anchored. In other words, it may be optimal for some of the
less informed to listen and not speak for a while as they build up their own
\textquotedblleft initial opinions\textquotedblright\ before joining the
public conversation.

Another important assumption of our model is that the underlying state
$\theta$ is fixed. In particular, there would be no reason to keep on seeding
in the initial opinions if the underlying state drifts. However it may still
be useful to use a FJ-type rule where the private seed is periodically updated
by each player to reflect the private signals about $\theta$ that each one accumulates.

Finally, our approach evaluates rules based on their fitness value. With a
continuum of states\ and opinions modeled as point-beliefs, averaging opinions
naturally has some fitness value. When there are few states and opinions take
the form of probabilistic beliefs, averaging beliefs or log-beliefs will
generally have poor (if not negative) fitness value (see for example
\cite{sobel14}). In this context, a promising FJ-like rule would consist in
linearly aggregating the \textit{initial change} in one's own log-belief
(induced by one's initial signal) with the perceived \textit{change} in a
composite neighbor's log-beliefs: such a rule accommodates the intuition that
\textit{belief changes} potentially reveal information, and through
appropriate weighting of one's own versus other's changes, it enables each
player to deal with situations where initial belief updates are driven by
interpretation errors (one then needs to filter out interpretation errors and
averaging is good in these cases) and situations where independent information
needs to be aggregated (adding changes in log-beliefs across all players would
be called for). Furthermore, as in this paper, it allows beliefs to differ and
the anchoring on one's own initial information (i.e., the initial change in
one's own log-belief) can limit the damaging effects of cumulated processing errors.

\bibliographystyle{plainnat}

\section*{{\protect\Large Appendix A}}

\textbf{Notations. }Define $M$ and $\Gamma$ as the $N\times N$ diagonal
matrices where $M_{ii}=m_{i}$ and $\Gamma_{ii}=\gamma_{i}$. For any fixed
vectors of signals $x$ and systematic bias $\xi$, we let
\[
X=Mx+(I-M)\xi
\]
and, whenever $m_{i}>0$, we let $\widetilde{x}_{i}=x_{i}+\xi_{i}%
(1-m_{i})/m_{i}$ denote the modified initial opinion, and $\widetilde{x}%
=(\widetilde{x}_{i})_{i}$ the vector. Next define the matrix $B=I-\Gamma
+\Gamma(I-M)A$.

We shall say that $P$ is a \textit{probability matrix} if and only if
$\sum_{j}P_{ij}=1$ for all $i$. Note that $A$ is a probability matrix and
throughout, we assume that the power matrix $A^{k}$ only has strictly positive
elements for some $k$. Finally, we refer to $v(y)$ as the variance of
$y$.\smallskip\textbf{ \smallskip}

In the main text, we show that when $m_{i}>0$\textit{ }for all $i$, long-run
opinions are weighted averages of modified opinions $\widetilde{x}$. Lemma~A1
below (proved in Appendix B) generalizes this observation. Define $N^{0}$
$\varsubsetneq N$ as the set of $n_{0}$ agents following DG ($m_{i}=0$).
Denote by $\xi^{0}$ the vector of errors of these players. We
have:\textbf{\smallskip}

\textbf{Lemma A1. }\textit{Assume }$n_{0}<n$\textit{. Then }$y=P\widetilde{x}%
+Q\xi^{0}$\textit{ where }$P$\textit{ is a }$(n,n-n_{0})$\textit{-probability
matrix.}\textbf{\smallskip}

Proposition 3 is then obtained as an immediate corrolary of Lemma~5.
\textbf{\smallskip}

\textbf{Proof of Proposition 3}: From Lemma A1, $L_{i}=var(y)\geq\frac{1}%
{n}\min var(\widetilde{x}_{i})\geq\frac{(1-m)^{2}\varpi}{n\text{ }m^{2}%
}.\blacksquare$\textbf{\smallskip}

We now turn to the proof of our main Propositions.

\textbf{Proof of Proposition 4:} Assume $m>>0$ so $\widetilde{x}_{j}$ is
well-defined for all $j$.\footnote{Cases where some or all $m_{j}$ are $0$ can
be derived by taking limits as $Q^{i}$ remains well-defined.} For $j\neq i$
let $X_{j}=m_{j}\widetilde{x}_{j}+(1-m_{j})A_{ji}y_{i}$ and $c_{j}^{i}%
=m_{j}+(1-m_{j})A_{ji}$. (\ref{Eqyk}) can be written in matrix form to obtain,
by definition of $Q^{i}$, $y_{-i}=Q^{i}X$. Note that if $\widetilde{x}_{j}=1$
for all $j$ and $y_{i}=1$, then $y_{k}=1$ for all $k$, so $\sum_{j\neq
i}Q_{kj}^{i}c_{j}^{i}=1$ for all $k$, which implies
\begin{equation}
\sum_{j\neq i}Q_{kj}^{i}(1-m_{j})A_{ji}=1-\sum_{j\neq i}Q_{kj}^{i}m_{j},
\label{Eqp3}%
\end{equation}
and, since $Q^{i}\ $is a positive matrix,\footnote{\label{ftprop5}$Q^{i}%
=\sum_{n\geq0}((I-M^{i})(I-\alpha^{i})\widetilde{A}^{i})^{n}$ so $Q^{i}$ is
non-negative. If in addition, $m_{-i}<<1$, and since $A$ is connected, then
$Q^{i}>>0$.} $\sum_{j\neq i}Q_{kj}^{i}m_{j}\leq1$, so $\sum_{j\neq i}R_{j}%
^{i}m_{j}\leq1$. (\ref{Eqp3}) further implies%
\begin{equation}
y_{k}=\sum_{j\neq i}Q_{kj}^{i}m_{j}\widetilde{x}_{j}+(1-\sum_{j\neq i}%
Q_{kj}^{i}m_{j})y_{i}, \label{Eqp4}%
\end{equation}
thus characterizing the influence of $y_{i}$ on $k$'s opinion. In particular,
the smaller $\sum_{j\neq i}Q_{kj}^{i}m_{j}$ the larger the influence of $i$ on
$k$. Averaging over all neighbors of $i$, and taking into account the weight
$A_{ik}$ that $i$ puts on $k$, we obtain:%
\begin{equation}
y_{i}=m_{i}\widetilde{x}_{i}+(1-m_{i})(\sum_{j\neq i}R_{j}^{i}m_{j}%
\widetilde{x}_{j}+y_{i}(1-\sum_{j\neq i}R_{j}^{i}m_{j}) \label{Eqp4b}%
\end{equation}
which, since $m_{j}\widetilde{x}_{j}=m_{j}x_{j}+(1-m_{j})\xi_{j}$ and
$h_{i}=1/\sum_{j\neq i}R_{j}^{i}m_{j}$, gives the desired Expressions
(\ref{Propyi}) for $y_{i}$, $\widehat{x}_{i}$, $p_{i}$ and $\widehat{\xi}_{i}%
$.$\blacksquare\medskip$

\textbf{Proof of Proposition 5}: Assume $m_{i}>0$ and apply Proposition~4,
taking the limit where all $m_{j}$ tend to $0$. For $A$ given, $Q^{i}$ and
$R^{i}$ are uniformly bounded (with a well-defined limit when all $m_{j}$
tends to $0$), and $(1-p_{i})\widehat{x}_{i}$ tends to $0$, which concludes
the proof.$\blacksquare$ $\medskip$

\textbf{Proof of Proposition 6. }Player $i$ optimally sets $p_{i}$ such that
$\frac{p_{i}}{1-p_{i}}=\frac{v(\widehat{x}_{i}+\widehat{\xi}_{i})}{v(x_{i}%
)}=\frac{W_{i}}{\sigma_{i}^{2}}$. Substituting $p_{i}$, we get the desired
expression for $L_{i}$.$\blacksquare\medskip$

\textbf{Proof of Result 1:} There are two parts in this proof. We first prove
that the $m_{i}^{\prime}s$ cannot be positive. Next we show that the
equilibrium outcome must be efficient. Recall $\pi^{\ast}=\arg\min_{\pi}%
v(\sum_{k}\pi_{k}x_{k})$ is the efficient weighting of seeds and $v^{\ast
}\equiv v(\pi^{\ast}.x)$. Also let $r_{i}=1/h_{i}$.

Assume by contradiction that $m_{j}>0$. Then (\ref{Propyi}) implies that
$m_{i}>0$ for all $i$, so $m>>0$. Next, from (\ref{Eqp4b}), and letting
$r_{i}=1/h_{i}$, we obtain $\widehat{y}_{i}=r_{i}\widehat{x}_{i}%
+(1-r_{i})y_{i}$, hence substituting $y_{i}$,
\begin{equation}
\widehat{y}_{i}=(1-r_{i})p_{i}x_{i}+(1-(1-r_{i})p_{i})\widehat{x}_{i}.
\label{Eqz}%
\end{equation}
So both $\widehat{y}_{i}$ and $y_{i}$ are weighted average between $x_{i}$ and
$\widehat{x}_{i}$, and since $m>>0$, $r_{i}\in(0,1)$, the weights are
different. Since $i$ optimally weighs $x_{i}$ and $\widehat{x}_{i}$ (using
$p_{i}$ on $x_{i}$), the weight $(1-r_{i})p_{i}$ is suboptimal\ so
\begin{equation}
v(y_{i})<v(\widehat{y}_{i})\leq\max_{j\neq i}v(y_{j}), \label{Eqzv}%
\end{equation}
where the second inequality follows from $\widehat{y}_{i}$ being an average of
the $y_{j}$'s. Since (\ref{Eqzv}) cannot be true for all $i$, we get a
contradiction. The equilibrium must thus be DG.

Consider now a DG equilibrium. Call $\pi=(\pi_{i})_{i}$ the weights on seeds
induced by $\gamma$ and $A$, $\widehat{\pi}^{i}$ the relative weights on
$k\neq i$, and $\widehat{x}_{i}=\widehat{\pi}^{i}.x_{-i}$. We have $y_{i}%
=\pi_{i}x_{i}+(1-\pi_{i})\widehat{x}_{i}$, and modifying $\gamma_{i}$ allows
the agent to modify $\pi_{i}$ without affecting $\widehat{x}_{i}$ (player $i$
increases $\pi_{i}$ by decreasing $\gamma_{i}$). Therefore the optimal choice
$\pi_{i}$ satisfies
\[
\frac{\pi_{i}}{1-\pi_{i}}=\frac{v(\widehat{x}_{i})}{\sigma_{i}^{2}}%
\]
Let $W_{i}^{\ast}=\min_{q}v(q.x_{-i})$. Since optimal weighting of all seeds
requires optimal weighting on seeds other than $i,$ we have:%
\[
\frac{\pi_{i}^{\ast}}{1-\pi_{i}^{\ast}}=\frac{W_{i}^{\ast}}{\sigma_{i}^{2}}%
\]
which implies
\begin{equation}
\pi_{i}=\pi_{i}^{\ast}+\frac{(1-p_{i})(1-\pi_{i}^{\ast})}{\sigma_{i}^{2}%
}(v(\widehat{x}_{i})-W_{i}^{\ast}) \label{Eqv}%
\end{equation}
Since all $\pi_{i}$ (and $\pi_{i}^{\ast})$ add up to one, one must have
$v(\widehat{x}_{i})-W_{i}^{\ast}\leq0$, hence information aggregation is
perfect.$\blacksquare$\smallskip

Before showing Result 2, we start with two Lemma that we also use to prove
Result 3:\smallskip

\textbf{Lemma A2:} \textit{For each }$j\neq i$\textit{, there exists }%
$\mu_{ji}$\textit{ and a probability vector }$C^{ji}\in\Delta_{N-1}$\textit{,
each independent of }$m_{i}$\textit{, such that}
\begin{equation}
y_{j}=(1-\mu_{ji})C^{ji}\widetilde{x}_{-i}+\mu_{ji}y_{i} \label{yk}%
\end{equation}

\textbf{Proof:} This immediately follows from Expression (\ref{Eqp3}) in the
proof of Proposition 4.$\blacksquare$\smallskip

\textbf{Lemma A3:} if\textit{ }$\frac{\partial L_{i}}{\partial m_{i}}\leq
0$\textit{, then }$\frac{\partial L_{j}}{\partial m_{i}}<0$\textit{ for all
}$j$\textit{.} \smallskip\smallskip

\textbf{Proof:} Since $\mu_{ji}$ and $C^{ji}$ are independent of $m_{i}$, we
obtain:%
\[
\frac{\partial L_{j}}{\partial m_{i}}=(\mu_{ji})^{2}\frac{\partial L_{i}%
}{\partial m_{i}}+\mu_{ji}(1-\mu_{ji})\sum_{k\neq i}C_{k}^{ji}\frac{\partial
Cov(\widetilde{x}_{k}y_{i})}{\partial m_{i}}%
\]
We substitute $y_{i}=p_{i}x_{i}+(1-p_{i})(\widehat{x}_{i}+\widehat{\xi}_{i})$
(see (\ref{Propyi})). Since $\widetilde{x}_{k}$ and $x_{i}$ are independent,
and since $\widehat{x}_{i},\widetilde{x}_{k}$ and $\widehat{\xi}_{i}$ do not
depend on $m_{i}$, we get
\[
\frac{\partial L_{j}}{\partial m_{i}}=(\mu_{ji})^{2}\frac{\partial L_{i}%
}{\partial m_{i}}-\mu_{ji}(1-\mu_{ji})\frac{\partial p_{i}}{\partial m_{i}%
}\sum_{k\neq i}C_{k}^{ji}Cov(x_{k}\widehat{x}_{i}+\widetilde{x}_{k}%
\widehat{\xi}_{i})
\]
The terms$\frac{\partial p_{i}}{\partial m_{i}}$ and $Cov(x_{k}\widehat{x}%
_{i})$ are positive, and so are the terms $Cov(\widetilde{x}_{k}\widehat{\xi
}_{i})$ when persistent errors are independent or positively correlated. The
sum on the right side is thus positive (and the effect is amplified with
errors), which proves Lemma~7.$\blacksquare$\smallskip

\textbf{Proof of Result 2: }Let $\underline{m}=\varpi/(1+\varpi)$. We show
that DG and all strategies $m_{i}<\underline{m}$ are dominated by
$\underline{m}$.

Assume first that all other players use $DG$. Then by Proposition~5, $L_{i}$
decreases strictly with $m_{i}$. Now assume that at least one player $j$
chooses $m_{j}>0$. Then $L_{i}=p_{i}^{2}+(1-p_{i})^{2}v(\widehat{x}%
_{i}+\widehat{\xi}_{i})$. Whether persistent errors are independent or fully
correlated, the variance of $\widehat{\xi}_{i}$ is at least equal to
$h_{i}^{2}\varpi$, which implies that $L_{i}$ strictly decreases with $p_{i}$
when $\frac{p_{i}}{1-p_{i}}<h_{i}^{2}\varpi$, hence also with $m_{i}$ when
$\frac{m_{i}}{1-m_{i}}<h_{i}\varpi$, and from Lemma~A2, we conclude that
$L_{j}$ decreases as well (on this range of $m_{i}$).$\blacksquare$\smallskip

\textbf{Proof of Result 3:}

\textbf{Step 1: lowerbounds\ on }$\overline{m}^{i}\equiv\max_{j\neq i}m_{j}$.

With transmission errors, optimal weighting of $x_{i}$ and $\widehat{x}_{i}$
implies%
\begin{equation}
\frac{p_{i}}{1-p_{i}}=\frac{v(\widehat{x}_{i})+v(\widehat{\xi}_{i})}%
{\sigma_{i}^{2}}%
\end{equation}
and (\ref{Eqv}) becomes
\begin{equation}
p_{i}=\pi_{i}^{\ast}+\frac{(1-p_{i})(1-\pi_{i}^{\ast})}{\sigma_{i}^{2}%
}(v(\widehat{x}_{i})-v_{i}^{\ast}+v(\widehat{\xi}_{i})) \label{Eqv2}%
\end{equation}
The weight $p_{i}$ is thus necessarily above the efficient level $\pi
_{i}^{\ast}$, and there are now two motives for doing that: inefficient
aggregation by others, and the cumulated error term $\widehat{\xi}_{i}$.

While (\ref{Eqv2}) implies a lower bound on $p_{i}$, as (\ref{Eqv}) did, there
is a major difference here with the no noise case where DG is used by all:
$p_{i}$ is the weight that $i$ puts on own seed, but since there is no
consensus, the sum $\sum_{i}p_{i}$ is not constrained to be below 1.
Nevertheless, when all $m$ are small, $\sum_{i}p_{i}=1+O(m)$ is close to 1,
and this allows us to bound $v(\widehat{\xi}_{i})$ (and the difference
$v(\widehat{x}_{i})-v_{i}^{\ast}$), as we now explain.

From Proposition 4, each opinion $y_{i}$ may be written as $y_{i}%
=P^{i}x+(1-P_{i}^{i})\widehat{\xi}_{i}$, where $P^{i}$ is a weighting vector
(such that $P_{i}^{i}=p_{i}$). (\ref{Eqp4}) implies that when all $m$ are
small, the vectors $P^{i}$ must be close to one another: seeds must be
weighted in almost the same way, and differences in opinions are mostly driven
by the terms $\widehat{\xi}_{i}$. Specifically, let $\overline{m}^{i}%
=\max_{j\neq i}m_{j}$. (\ref{Eqp4}) implies that for all $k\neq i$,
\[
p_{k}=P_{k}^{k}\leq P_{k}^{i}+c\overline{m}^{i}%
\]
for some constant $c$ independent of $m$ and $k$. Since $P_{kk}=p_{k}\geq
\pi_{k}^{\ast}$, adding these inequalities yield%
\begin{equation}
1-p_{i}=\sum_{k\neq i}P_{k}^{i}\geq\sum_{k\neq i}p_{k}-Kc\overline{m}^{i}%
\geq1-\pi_{i}^{\ast}-Kc\overline{m}^{i} \label{EqProp2}%
\end{equation}
which, combined with (\ref{Eqv2}) yields, for some constant $d$,%
\begin{equation}
\overline{m}^{i}\geq d(v(\widehat{x}_{i})-v_{i}^{\ast}+\frac{\varpi
}{(\overline{m}^{i})^{2}}). \label{EqProp2b}%
\end{equation}
Since $var(\widehat{x}_{i})-v_{i}^{\ast}\geq0$, this implies $\overline{m}%
^{i}\geq(d\varpi)^{1/3}$, which further implies that the variance
$v(\widehat{\xi}_{i})$ is at most comparable to $\varpi^{1/3}$.\smallskip

\textbf{Step 2: upperbounds on }$\overline{m}^{i}$\textbf{.} Let $r_{i}%
=\sum_{j\neq i}R_{j}m_{j}$ and $\widehat{y}_{i}=\sum_{k\neq i}A_{ik}y_{k}$.
With transmission errors, we obtain:%
\[
\widehat{y}_{i}=(1-r_{i})p_{i}x_{i}+(1-(1-r_{i})p_{i})(\widehat{x}%
_{i}+\widehat{\xi}_{i})+\overline{\xi}_{i}%
\]
where $\overline{\xi}_{i}=-p\xi_{i}+(1-p_{i})\sum_{j\neq i}R_{j}(1-m_{j}%
)\xi_{j}$. Since $p_{i}$ is set optimally by $i$, we have:%
\[
v(\widehat{y}_{i})-v(y_{i})\geq(r_{i}p_{i})^{2}(\sigma_{i}^{2}+v(\widehat{x}%
_{i})+v(\widehat{\xi}_{i}))-E\overline{\xi}_{i}-(1-p_{i})E\overline{\xi}%
_{i}\widehat{\xi}_{i}\geq cr_{i}^{2}-\frac{d\varpi}{r_{i}}%
\]
for some constant $c$ and $d$ (independent of $\varpi$ and $m$). Since
$v(\widehat{y}_{i})\leq\max v(y_{k})$, the right-hand side cannot be positive
for all $i$, so $r_{i_{0}}\leq(d\varpi/c)^{1/3}$ for some $i_{0}$. From step
$1$, we conclude that $\overline{m}^{i_{0}}$ and all $m_{j}$ with $j\neq
i_{0}$ are $O(\varpi^{1/3})$, and that $m_{i_{0}}$ is thus \textit{at least}
$O(\varpi^{1/3})$.

It only remains to check that $m_{i_{0}}$ cannot be large. From (\ref{EqProp2}%
), $p_{i_{0}}\leq\pi_{i_{0}}^{\ast}+O(\varpi^{1/3})$, and since $p_{i_{0}}%
\geq\frac{1}{1+r_{i_{0}}/m_{i_{0}}}$, we conclude that all $m_{i}$ (and thus
$\overline{m}^{i}$) are $O(\varpi^{1/3})$, which further implies that all
variances $v(\widehat{\xi}_{i})$ are $O(\varpi^{1/3})$.

These variances imply that $Ey_{i}^{2}-v^{\ast}$ is at least $O(\varpi^{1/3}%
)$. $Ey_{i}^{2}$ also rises because of inefficient weighting of seeds, but the
loss is of the order of $(p_{i}-\pi_{i}^{\ast})^{2}$, that is, $O(\varpi
^{2/3})$, a significantly lower loss.$\blacksquare$\smallskip

\textbf{Proof of Result 4:} this follows from Lemma 7 since at equilibrium
$\frac{\partial L_{i}}{\partial m_{i}}=0$. $\blacksquare\smallskip$

\textbf{Proof of Expression~(\ref{pioverpk}).} Call $p_{j}^{i}$ the weight
that $i$ puts on $j$ and $\overline{R}_{i}$ the limit of $R_{i}$ when $m_{-i}$
tends to $0$. It follows from Proposition~4 when all $m$ are small,
$(p_{j}^{i}/m_{j})/(p_{i}/m_{i})\simeq\overline{R}_{ij}$. To compute
$\overline{R}_{ij}$, consider the case where $m_{i}=m$ for all $i$. Then
$y=m\widetilde{x}+(1-m)A\widetilde{x}=\sum m(1-m)^{k}A^{k}\widetilde{x}$.
Since all lines of $A^{k}$ are close to $\rho$ when $k$ is large enough,
$y_{i}\simeq\rho\widetilde{x}$ for all $i$, so $\overline{R}_{ij}=\rho
_{j}/\rho_{i}$.\smallskip

\textbf{Proof of (generalized) Result~5}: Rather than assuming that the
central player is DG, we consider here a central player who uses her seed
$x_{0}$ optimally to minimize the loss $v(\overline{y})$, given $m$. We have
$\overline{y}=(1-m)y_{0}$ and $y_{0}=m_{0}x_{0}+(1-m_{0})(\overline{y}+\xi
_{0})$. This gives $\overline{y}=(1-m)(p_{0}x_{0}+(1-p_{0})\frac{\xi_{0}}{m})$
where the central player controls $p_{0}$. The variance $v(\overline{y})$ is
minimized for $\frac{p_{0}}{1-p_{0}}=\frac{\varpi_{0}}{m^{2}}$, and we get
$v(\overline{y})=(1-m)^{2}\frac{\varpi_{0}/m^{2}}{1+\varpi_{0}/m^{2}}$. So
long as $m>>(\varpi_{0})^{1/2}$, we obtain $D\simeq\frac{4\varpi_{0}}{d}$ as
for the DG case. Note that when $m\leq O(\varpi_{0})^{1/2}$, cumulated errors
are potentially huge and the (benevolent) central player mitigates them by
choosing a large $m_{0}$: since she is benevolent, the loss cannot exceed 1
(the variance of her own seed).$\blacksquare$\smallskip\smallskip

We conclude this Appendix with network comparisons. We derive Proposition~A0
(see below), on which the discussion in the main text is based, and which we
prove in Appendix~B.\textbf{ }

To facilitate network comparisons, we assume initial signals of identical
precision ($\sigma_{i}^{2}=1$), so that the efficient weighting of signals is
$\pi_{i}^{\ast}=1/n$ and $\underline{W}_{i}^{\ast}\equiv var(\pi_{-i}^{\ast
}.x_{-i})=1/(n-1)$. All players are subject to a processing error $\xi_{i}$,
with same variance $\varpi$. From Proposition~6, player $i$'s incentives
yields
\begin{equation}
\frac{m_{i}}{1-m_{i}}=W_{i}/h_{i} \label{cond}%
\end{equation}
where $W_{i}=var(\widehat{x}_{i})+var(\widehat{\xi}_{i})$. Both $h_{i}$ and
$W_{i}$ depend only $m_{-i}$ and the structure of the network, and the
equilibrium values $m_{i}^{\ast}$ are obtained by simultaneously solving these
equations. Given this equilibrium values, we can then compute $W_{i}^{\ast}$
hence (by Proposition~6), the equilibrium loss $L_{i}^{\ast}$. To measure how
losses $L_{i}$ departs from the minimum loss $\underline{L}_{i}^{\ast}$, we
define
\[
\widehat{\Delta}_{i}\equiv W_{i}-\underline{W}_{i}^{\ast}%
\]
which characterizes the size of the inefficiency resulting from the
inefficient aggregation of others' signals and cumulated errors. Defining%
\[
\rho_{i}\equiv\frac{p_{i}}{1-p_{i}}-\frac{1}{n-1}=\frac{m_{i}h_{i}}{1-m_{i}%
}-\frac{1}{n-1}%
\]
the equilibrium condition can thus be written,%
\[
\rho_{i}=\widehat{\Delta}_{i}%
\]
which has the following economic interpretation: the relative weight on
$x_{i}$ (relative to other signals) should exceed the efficient weighting by
$\widehat{\Delta}_{i}$.

We compare three $n$-player networks: the \textit{complete network}, where
each player is connected to all others; the \textit{directed circle}, where
information transmission is directed and one-sided (player $i$ communicates to
player $i-1$, who communicates to $i-2$, and so on -- player $0$ is player
$n$) and the \textit{star network} which consists of $n-1$ peripheral players
labelled $k=1,...n-1$ and a central player, labelled $0$, who aggregates the
opinions of the peripheral players.

For each network, we characterize $h_{i}$, $\widehat{x}_{i}$ and
$\widehat{\xi}_{i}$ (hence $\rho_{i}$ and $\widehat{\Delta}_{i}$) indicating a
superscript $c$ for the complete network$,$ $d$ for the directed circle and
$s$ for the star network. We next solve for equilibrium, focusing on the limit
cases where $\varpi$ is small (for a fixed $n$) and where for a fixed $\varpi$
small, $n$ gets large. For the complete network and the directed circle, we
solve for a symmetric equilibrium. For the star network, we solve for an
equilibrium where all peripheral players use the same weight $m$, and the
central player, labelled player $0$, uses $m_{0}$. We obtain:\smallskip

\textbf{Proposition A0}: \textit{For fixed }$n\geq3$\textit{ and small
}$\varpi$\textit{, }$\widehat{\Delta}_{d}^{\ast}<\widehat{\Delta}_{c}^{\ast
}<\widehat{\Delta}_{s}^{\ast}$\textit{. For fixed small }$\varpi$\textit{, at
the large }$n$\textit{ limit, }$\widehat{\Delta}_{c}^{\ast}<\widehat{\Delta
}_{d}^{\ast}<\widehat{\Delta}_{s}^{\ast}$\textit{. These comparisons hold
whether errors are independent or correlated. Furthermore, for the star
network, }$m_{0}^{\ast}/m^{\ast}\simeq\frac{1}{n-1}$ \textit{for fixed }%
$n$\textit{ and small }$\varpi$\textit{, and }$m_{0}^{\ast}/m^{\ast}%
\leq(2\varpi)^{1/3}$\textit{ at the large }$n$\textit{ limit.\smallskip}

\newpage

\section*{Appendix B (for on-line publication)}

\subsection*{B.1 Proposition 2}

We first prove that the matrix $H\equiv\sum_{k\geq0}B^{k}$ is well-defined
(Lemma B1 and B2), and obtain Proposition 2 as a Corollary. \smallskip

\textbf{Lemma B1:} \textit{Consider any non-negative matrix }$C=(c_{ij})_{ij}%
$\textit{ such that }$\mu=\min_{i}(1-\sum_{j}c_{ij})>0$\textit{. Then }%
$I-C$\textit{ has an inverse }$H\equiv\sum_{k\geq0}C^{k}$\textit{, and for any
}$X^{0}$\textit{ and }$Y^{0}$\textit{, }$Y^{t}=X^{0}+CY^{t-1}$\textit{
converges to }$HX^{0}$\textit{.}\smallskip

\textbf{Lemma B2:} \textit{If }$m_{i_{0}}>0$\textit{, then for }$K$\textit{
large enough, }$C=B^{K}$\textit{ satisfies the condition of Lemma 1, and
}$I-B$\textit{ has an inverse. }\smallskip

\textbf{Proof of Proposition 2: }We just need to check that $y^{t}$ converges.
We iteratively substitute in (\ref{Ybar}) to get:
\[
y^{t}=X^{0}+Cy^{t-K}%
\]
where $X^{0}=D\Gamma X$ with $D\equiv I+B+...+B^{K-1}$, and $C=B^{K}$. By
Lemma 2, Lemma 1 applies to $C$, so convergence of $y^{t}$ is
ensured.$\blacksquare$\smallskip

\textbf{Proof of Lemma B1:} Consider the matrix $H^{t}=(h_{ij}^{t})_{ij}$
defined recursively by $H^{0}=I$ and $H^{t}=I+CH^{t-1}$. Let $z^{t}=\max
_{ij}|h_{ij}^{t}-h_{ij}^{t-1}|$. We have $z^{t}\leq(1-\mu)z^{t-1}$, implying
that $H^{t}$ has a well-defined limit $H$, which satisfies $H\equiv\sum
_{k\geq0}C^{k}$. By construction, $(I-C)H=H(I-C)=I$, so $H=(I-C)^{-1}$.
Similarly, defining $z^{t}=\max_{i}\left\vert Y_{i}^{t}-Y_{i}^{t-1}\right\vert
$, we obtain that $Y^{t}$ has a limit $Y$ which satisfies $(I-C)Y=X^{0}$,
implying $Y=HX^{0}$.$\blacksquare\smallskip\smallskip$

Before turning to the proof of Lemma~B2, we define \textit{sequences},
\textit{paths} and probabilities over paths associated with a probability
matrix $A=(A_{ij})_{ij}$. For any sequence $q=(i_{1},...,i_{K})$, we let
$\pi^{A}(q)\equiv%
{\textstyle\prod\nolimits_{k=1}^{K-1}}
A_{i_{k},i_{k+1}}$, and for any set of sequences $Q$, we abuse notations and
let $\pi^{A}(Q)=\sum_{q\in Q}\pi^{A}(q)$. We define a \textit{path} as a
sequence $q$ for which $\pi^{A}(q)>0$.\smallskip

Denote by $Q_{i,j}^{K}$ the set of paths of length $K$ from $i$ to $j$, and
$Q_{i}^{K}$ the set of paths of length $K$ that start from $i$. $Q_{i}%
^{K}=\cup_{j}Q_{i,j}^{K}$ and by construction, for any $i$, $j$%
\begin{equation}
A_{ij}^{K}\equiv\pi^{A}(Q_{i,j}^{K})\text{ and }\sum_{j\in N}A_{ij}^{K}%
=\pi^{A}(Q_{i}^{K})=1 \label{eqP}%
\end{equation}
where $A^{K}$ is the $K^{th}$ power of matrix $A$.$\smallskip\smallskip$

\textbf{Proof of Lemma B2:} We consider $A$ connected, that is, such that
$A_{ij}^{k}>0$ for all $i,j$, and consider $K\geq2k$. Call $Q_{i}^{K,i_{0}%
}\subset Q_{i}^{K}$ the set of paths of length $K$ that start from $i$ (to
some $j$) and go through $i_{0}$. For any such path, $\pi^{B}(q)\leq
(1-\underline{\gamma}m_{i_{0}})\pi^{A}(q)$. This implies
\[
\sum_{j}C_{ij}\equiv\pi^{B}(Q_{i}^{K})\leq(1-\underline{\gamma}m_{i_{0}}%
)\pi^{A}(Q_{i}^{K,i_{0}})+\pi^{A}(Q_{i}^{K}\backslash Q_{i}^{K,i_{0}})<1
\]
where the last inequality follows from (\ref{eqP}) and $Q_{i}^{K,i_{0}}$ non
empty for $K\geq2k$. This implies that $C$ satisfies the condition of Lemma 1,
hence $I-C$ has an inverse. Let $D\equiv I+B+...+B^{K-1}$ and $H=(I-C)^{-1}D$.
We have
\[
\sum_{k\geq0}B^{k}=\sum_{k\geq0}C^{k}D=H,
\]
so $H(I-B)=(I-B)H=I$ and $I-B$ also has an inverse.$\smallskip\smallskip$

\textbf{Proof of Lemma A1}: Using the recursive equation $y=X+(I-M)Ay,$ and
$X_{i}=m_{i}\widetilde{x}_{i}$ for $i\notin N^{0}$ and $X_{i}=\xi_{i}^{0}$ for
$i\in N^{0}$, we define recursively the $(n,n-n_{0})$ and $(n,n_{0})$ matrices
$P^{t}$ and $Q^{t}$ as follows: for $i\notin N^{0}$, we let $P_{i}^{t}%
=m_{i}+(1-m_{i})A_{i}P^{t-1}$ and $Q_{i}^{t}=(1-m_{i})A_{i}Q^{t-1}$, and for
$i\in N^{0}$, $P_{i}^{t}=A_{i}P^{t-1}$ and $Q_{i}^{t}=I+A_{i}Q^{t-1}$. Also we
let $P_{ii}^{1}=1$ for $i\notin N^{0}$, and all other $P_{ij}^{1}$ and all
$Q_{ij}^{1}$ equal to $0$. By construction, $y=P\widetilde{x}+Q\xi_{0}$ where
$P$ and $Q$ are the limit of $P^{t}$ and $Q^{t}$ respectively. Besides, by
induction on $t$, each $P^{t}$ is a probability matrix, hence so is the limit
$P.$
$\smallskip\smallskip$

\subsection*{B.2 Network Comparisons}

Before proving Proposition\ A0, we gather a number of preliminary results,
deriving $h_{i}$, $\widehat{x}_{i}$ and $\widehat{\xi}_{i}$ for each network.
For the complete network and the directed circle, we analyze a situation where
all players but $i$ use the same weight $m$. We denote by $\overline{x}_{-i}$
(respectively $\overline{\xi}_{-i}$) the average seed (respectively error) of
players other than $i$, and for any $z_{-i}$, let $\psi(z_{-i})=\sum
_{k=1}^{n-1}(1-m)^{k-1}z_{i+k}/\sum_{k=1}^{n-1}(1-m)^{k-1}$ be the weighted
average over $z_{i+k}$'s where the weight of the $k-$step neighbor is
diminished by a factor $(1-m)^{k-1}$. Simple computations show:

\textbf{Lemma B3}: \textit{For the complete network, }$h_{i}^{c}=1+\frac
{1-m}{m(n-1)}$\textit{, }$\widehat{x}_{i}^{c}=\overline{x}_{-i}$, \textit{and
}$\widehat{\xi}_{i}^{c}=h_{i}^{c}\xi_{i}+\overline{\xi}_{-i}\frac{1-m}{m}%
$\textit{. For the directed circle, }$h_{i}^{d}=\frac{1}{1-(1-m)^{n-1}}%
(<h_{i}^{c})$\textit{ and }$\widehat{x}_{i}^{d}=\psi(x_{-i})$\textit{ and
}$\widehat{\xi}_{i}^{d}=h_{i}^{d}\xi_{i}+\psi(\xi_{-i})\frac{1-m}{m}$.

We next use Lemma~B3 to compute $\rho_{i}$ and $v(\widehat{x}_{i})$ for each
network:\smallskip

\textbf{Lemma~B4:} \textbf{ }\textit{For a fixed }$n$ and \textit{small} $m,$
and for i.i.d random variables, $v(\psi(x_{-i}))=v(x_{i})(\frac{1}{n-1}%
+cm^{2})$ where $c=\frac{n(n-2)}{12(n-1)}$. \textit{At the large }$n$\textit{
limit, }$v(\psi(x_{-i}))=v(x_{i})\frac{m}{2}$\textit{. }

\textbf{Lemma~B5:} \textit{For fixed }$n$, \textit{small }$m$\textit{ and
}$m_{i}=m$, $\rho_{i}^{c}=m$ and $\rho_{i}^{d}=\mu m$ \textit{where }%
$\mu=\frac{n}{2(n-1)}$. For fixed small $m$ and $m_{i}=m$, at the large $n$
limit, $\rho_{i}^{c}\simeq\rho_{i}^{d}\simeq m$. \textit{ }\smallskip

There are thus two key differences between the directed circle and the
complete network. At the large $n$ limit, the essential difference is that the
direct circle performs poorer aggregation, which will imply (see below) that
$\widehat{\Delta}_{d}^{\ast}>\widehat{\Delta}_{c}^{\ast}$. For a fixed $n$
however, the effect of poorer information aggregation is second order in $m$,
so the essential difference is $\rho_{i}^{c}>\rho_{i}^{d}$, which will imply
$m_{d}^{\ast}>m_{c}^{c}$ (i.e. a stronger incentive to raise $m_{i}$), hence
$\widehat{\Delta}_{d}^{\ast}<\widehat{\Delta}_{c}^{\ast}$.

Regarding the star network, the resolution has to separate the analysis of the
central and peripheral players.

\textbf{Lemma~B6}: \textit{When all peripheral players use }$m$\textit{, we
have }$h_{0}^{s}=\frac{1}{m}$\textit{, }$\widehat{x}_{0}^{s}=\overline{x}%
$\textit{ and }$\widehat{\xi}_{0}^{s}=\frac{\xi_{0}+(1-m)\overline{\xi}}{m}%
$\textit{. When the central player uses }$m_{0}$\textit{ and other peripheral
players use }$m$\textit{, set }$\rho_{0}\equiv\frac{m_{0}}{(1-m_{0})m}%
-\frac{1}{n-1}$\textit{ and }$q_{0}=\frac{1/(n-1)+\rho_{0}}{1+\rho_{0}}%
$\textit{. We have }$\widehat{x}_{i}^{s}=q_{0}x_{0}+(1-q_{0})\overline{x}%
_{-i}$\textit{, }$h_{i}^{s}=1+\frac{1}{m(n-1)(1+\rho_{0})}$\textit{ and
}$\widehat{\xi}_{i}^{s}=\xi_{i}(1+\frac{1}{(n-1)(1+\rho_{0})})+\frac
{1}{(1+\rho_{0})}\widehat{\xi}_{0}$. \smallskip

Recalling that $\rho_{0}^{s}\equiv\frac{m_{0}h_{0}^{s}}{(1-m_{0})}-\frac
{1}{n-1}$, Lemma B6 implies that in equilibrium
\begin{equation}
\rho_{0}^{s}=\widehat{\varpi}_{0}^{s}\equiv v(\widehat{\xi}_{0}^{s})
\label{pho0}%
\end{equation}
The above equation determines $m_{0}$ as a function of $m$, and for a fixed
$m$ and small $\widehat{\varpi}_{0}^{s},$ we must thus have $m_{0}\simeq
\frac{m}{n-1}$. Intuitively, the central player's opinion influences (many)
peripheral players, so for information aggregation purposes, the central
player should compensate for that influence by setting a smaller $m_{0}$
compared to $m$. Furthermore, at the large $n$ limit,\ $m_{0}\simeq
m\widehat{\varpi}_{0}^{s}$, so when $\widehat{\varpi}_{0}^{s}$ is small (which
will be true in equilibrium when $\varpi$ is small), her behavior becomes
close to that of a DG player.

Lemma B6 tells us that when $\rho_{0}>0$, the aggregation of seeds is
distorted (i.e, $v(\widehat{x}_{i}^{s})>\frac{1}{n-1})$ and potentially,
$h_{i}^{s}<h_{i}^{c}$. However in equilibrium, the incentive condition
(\ref{pho0}) of the central player implies when $\widehat{\varpi}_{0}^{s}$ is
small (which will be true in equilibrium when $\varpi$ is small), the
distortion is negligible and $h_{i}^{s}\simeq h_{i}^{c}$. So losses
essentially come from the cumulated error terms (i.e., $\widehat{\varpi}%
_{i}^{s}\equiv var(\widehat{\xi}_{i}^{s})$), which are higher for the star
network compared to both other networks. We turn to the detailed
proof.\smallskip

\textbf{Proof of Proposition A0. }

(i) \textit{For the complete network}, at $m_{i}=m,$ $\rho_{i}^{c}\simeq m$,
so the equilibrium condition gives $m=v(\widehat{\xi}_{i}^{c})$. Using
Lemma~B3, for independent errors, $v(\widehat{\xi}_{i}^{c})\simeq\frac
{1}{m^{2}}(\frac{1}{(n-1)^{2}}+\frac{1}{n-1})$, while for correlated errors,
$v(\widehat{\xi}_{i}^{c})=\frac{1}{m^{2}}(\frac{1}{n-1}+1)^{2}$, from which we
derive $m_{c}^{\ast}$ in each case. For fixed $n$, with independent errors,
$\widehat{\Delta}_{c}^{\ast}\simeq m_{c}^{\ast}\simeq\varpi^{1/3}(\frac
{n}{(n-1)^{2}})^{1/3}$, while with perfectly correlated errors
$\widehat{\Delta}_{c}^{\ast}\simeq m_{c}^{\ast}\simeq\varpi^{1/3}(\frac
{n}{n-1})^{2/3}$. For fix $\varpi$ small and large $n$, $\widehat{\Delta}%
_{c}^{\ast}\simeq m_{c}^{\ast}\simeq(\frac{\varpi}{n})^{1/3}$ for independent
errors, and $\widehat{\Delta}_{c}^{\ast}\simeq m_{c}^{\ast}\simeq\varpi^{1/3}$
for perfectly correlated errors. \smallskip

(ii) \textit{For the directed network}, using Lemma~B4, the equilibrium
condition now gives $\mu m\simeq v(\widehat{\xi}_{i}^{d})$ for fixed $n$, and
$m\simeq\frac{m}{2}+v(\widehat{\xi}_{i}^{d})$ for the large$~n$ limit. For
fixed $n$, $v(\widehat{\xi}_{i}^{d})\simeq v(\widehat{\xi}_{i}^{c})$ (by
Lemma$~$B4 for independent errors, and because $\psi(\xi_{-i}))=\overline{\xi
}_{-i}$ for correlated errors). It follows that $m_{d}^{\ast}=\mu^{-1/3}%
m_{c}^{\ast}$ and $\widehat{\Delta}_{d}^{\ast}=\mu m_{d}^{\ast}=\mu
^{2/3}\Delta_{c}^{\ast}$ in both cases. At the large~$n$ limit, Lemma$~$B4
implies $v(\psi(\xi_{-i}))=\frac{m}{2}\varpi$ for independent errors, so
$m_{d}^{\ast}\simeq\widehat{\Delta}_{d}^{\ast}\simeq\varpi^{1/2}%
>\widehat{\Delta}_{c}^{\ast}$. For correlated errors, $v(\widehat{\xi}_{i}%
^{d})=v(\widehat{\xi}_{i}^{c})$, so the equilibrium condition gives
$m_{d}^{\ast}\simeq2v(\widehat{\xi}_{d}^{\ast})\simeq(2\varpi)^{1/3}$ and
$\widehat{\Delta}_{d}^{\ast}=\frac{m_{d}^{\ast}}{2}+v(\widehat{\xi}_{d}^{\ast
})\simeq m_{d}^{\ast}\simeq2^{1/3}m_{c}^{\ast}>\widehat{\Delta}_{c}^{\ast}$.

Note that at the large $n$ limit, in contrast to full network where
inefficiencies are solely driven by cumulated errors, the cumulated errors and
the poor averaging of seeds \textit{equally contribute} to the overall
loss.\smallskip

(iii)\textit{ For the star network}, by Lemma~B6, the equilibrium condition
for the central player gives $\rho_{0}^{s}=\widehat{\varpi}_{0}^{s}$, and for
a peripheral player it gives, for small $m$,
\begin{equation}
\frac{h_{i}^{s}m}{1-m}-\frac{1}{n-1}\simeq v(\widehat{x}_{i}^{s})-\frac
{1}{n-1}+v(\widehat{\xi}_{i}^{s}) \label{hs}%
\end{equation}
Omitting terms of order $2$ in $\rho_{0}$ or $m$, we have $\frac{h_{i}^{s}%
m}{1-m}-\frac{1}{n-1}\simeq m+\frac{m-\rho_{0}^{s}}{n-1}$, $v(\widehat{x}%
_{i}^{s})-\frac{1}{n-1}\simeq0$, and $v(\widehat{\xi}_{i}^{s})\simeq
v(\widehat{\xi}_{0}^{s})=\rho_{0}^{s}$, so (\ref{hs}) implies
\begin{equation}
m\simeq\rho_{0}^{s}=\widehat{\varpi}_{0}^{s}. \label{hs2}%
\end{equation}
For low $\varpi$, we thus have $m\simeq O(\varpi^{1/3})$, justifying the
omission of terms of higher order. For independent errors, and letting
$\varpi_{0}=v(\xi_{0})$, we obtain $\widehat{\Delta}_{0}^{s}\simeq
\widehat{\varpi}_{0}^{s}\simeq(\varpi_{0}+\frac{\varpi}{n-1})/m^{2}$, so
(\ref{hs2})~implies $m_{s}^{\ast}\simeq(\varpi_{0}+\frac{\varpi}{n-1}%
)^{1/3}\simeq\widehat{\Delta}_{s}^{\ast}$. For correlated errors we get
$m_{s}^{\ast}\simeq(\varpi_{0}+\varpi)^{1/3}\simeq\widehat{\Delta}_{s}^{\ast}%
$. When $\varpi_{0}=\varpi$, then in both cases, $\widehat{\Delta}_{s}^{\ast
}>\max(\widehat{\Delta}_{d}^{\ast},$ $\widehat{\Delta}_{c}^{\ast})$.

Furthermore, we have already mentioned that $m_{0}^{\ast}\simeq\frac{m^{\ast}%
}{n-1}$ for fixed $n$, and that at the large $n$ limit, $m_{0}^{\ast}\simeq
m^{\ast}\widehat{\varpi}_{0}^{s}$, which concludes the proof of Proposition A0
since $\widehat{\varpi}_{0}^{s}\leq(2\varpi)^{1/3}$.$\blacksquare$

\bigskip

\textbf{Proof of Lemma B3 and B6:} For each network, we write the equations
determining long-run opinions. Through appropriate subsitutions, we derive
these opinions as a function of seeds and errors. For the complete network, we
use%
\begin{align*}
y_{i}  &  =m_{i}\widetilde{x}_{i}+(1-m_{i})\overline{y}_{-i}\text{ and}\\
\overline{y}_{-i}  &  =m\overline{\widetilde{x}}_{-i}+(1-m)(\frac{1}{n-1}%
y_{i}+\frac{n-2}{n-1}\overline{y}_{-i})
\end{align*}
where $\overline{\widetilde{x}}_{-i}$ (and $\overline{y}_{-i})$ refer to the
mean modified seed (and opinion) of all players but $i$. Note $\overline
{\widetilde{x}}_{-i}=\overline{x}_{-i}+\frac{1-m}{m}\overline{\xi}_{-i}$.

For the directed circle, we use $y_{1}=m_{1}\widetilde{x}_{1}+(1-m_{1})y_{2}$
and repeatedly substitute $y_{i}=m\widetilde{x}_{i}+(1-m_{i})y_{i+1}$ to
obtain:
\[
y_{1}=m_{1}\widetilde{x}_{1}+(1-m_{1})(\sum_{k=0}^{N-2}(1-m)^{k}%
m\widetilde{x}_{k+2}+(1-m)^{N-1}y_{1}%
\]
To prove that $h_{i}^{d}<h_{i}^{c}$ for all $m\in(0,1)$, observe that the
inequality holds for $m$ close to $0$ and that for any $m\in(0,1)$ that would
satisfy $h_{i}^{d}(m)=h_{i}^{c}(m)$, we would have $\frac{\partial h_{i}^{c}%
}{\partial m}(m)>\frac{\partial h_{i}^{d}}{\partial m}(m)$, in contradiction
with $h_{i}^{d}<h_{i}^{c}$ for $m$ close to $0$.

For the star network, we first determine $h_{0}$, $\widehat{x}_{0}$, and
$\widehat{\xi}_{0}$ using
\[
y_{0}=m_{0}\widetilde{x}_{0}+(1-m_{0})\overline{y}\text{ and }\overline
{y}=m\overline{\widetilde{x}}+(1-m)y_{0}%
\]
where $\overline{\widetilde{x}}$ (and $\overline{y})$ refer to the mean
modified seed (and opinion) of \textit{peripherical} players. Next, to
determine $h_{i}$, $\widehat{x}_{i}$, and $\widehat{\xi}_{i}$, we use
\begin{align*}
y_{0}  &  =m_{0}\widetilde{x}_{0}+(1-m_{0})(\frac{1}{n-1}y_{i}+(1-\frac
{1}{n-1})\overline{y}_{-i})\\
y_{i}  &  =m_{i}\widetilde{x}_{i}+(1-m_{i})y_{0}\\
\overline{y}_{-i}  &  =m\overline{\widetilde{x}}_{-i}+(1-m)y_{0}%
\end{align*}
where $\overline{\widetilde{x}}_{-i}$ (and $\overline{y}_{-i})$ refer to the
mean modified seed (and opinion) of all peripherical players but
$i$.$\blacksquare$\smallskip

\textbf{Proof of Lemma B4 and B5}: Let $r=1-(1-m)^{n-1}$. We have:%
\[
v(\psi(x_{-i}))/v(x_{i})=\frac{\sum_{k=0}^{n-2}(1-m)^{2k}}{(\sum_{k=0}%
^{n-2}(1-m)^{k})^{2}}=\frac{(1-(1-m)^{2(n-1)})m^{2}}{m(2-m)r^{2}}%
=\frac{(2-r)m}{r(2-m)}%
\]
At the large $n$ limit, $r=1$, hence the desired result. For fixed $n$,
compute $\Delta=v(\psi(x_{-i}))/v(x_{i})-\frac{1}{n-1}$ considering terms of
order up to 2 in $m$. We have $r=(n-1)m(1-\ell m)$ where $\ell=\frac{n-2}%
{2}(1-\frac{(n-3)}{3}m)$, from which we obtain
\[
(n-1)\Delta\simeq\frac{1-\frac{r}{2}}{(1-\frac{m}{2})(1-\ell m)}-1\simeq
(\ell+\frac{1}{2})m-\frac{\ell m^{2}}{2}-\frac{r}{2}\simeq m^{2}\frac
{n(n-2)}{12}%
\]
Regarding Lemma~B6, the first statement is immediate. Regarding the directed
circle, $h_{i}^{d}=1/r,$ so we have
\[
(n-1)h_{i}^{d}\frac{m}{1-m}-1\simeq\frac{1}{(1-\ell m)(1-m)}-1\simeq
(1+\ell)m\simeq\frac{n}{2}m\blacksquare
\]


\subsection*{B.3 The case with independent errors}

For any fixed $(x,\xi)$, we define the expected opinion at $t$, $\overline
{y}_{i}^{t}=Ey_{i}^{t}$ and the vector of expected opinions $\overline{y}%
^{t}=(\overline{y}_{i}^{t})_{i}$. We further define $\eta^{t}=y^{t}%
-\overline{y}^{t}$, $w_{ij}^{t}=E\eta_{i}^{t}\eta_{j}^{t}$ and the vector of
covariances $w^{t}=(w_{ij}^{t})_{ij}$.

We define the $N^{2}$ vector $\Lambda$ with $\Lambda_{ij}=0$ if $i\neq j$,
$\Lambda_{ii}=(\gamma_{i}(1-m_{i}))^{2}\varpi^{0}$ and $\overline{B}$ the
($N^{2}\times N^{2})$ matrix where $\overline{B}_{ij}$ is the row vector
$(\overline{B}_{ij,hk})_{hk}$ with $\overline{B}_{ij,hk}=B_{ih}B_{jk}$.

For any fixed $(x,\xi)$, we define the expected opinion at $t$, $\overline
{y}_{i}^{t}=Ey_{i}^{t}$ and the vector of expected opinions $\overline{y}%
^{t}=(\overline{y}_{i}^{t})_{i}$. We further define $\eta^{t}=y^{t}%
-\overline{y}^{t}$, $w_{ij}^{t}=E\eta_{i}^{t}\eta_{j}^{t}$ and the vector of
covariances $w^{t}=(w_{ij}^{t})_{ij}$.

The evolution of opinions and expected opinions (given $x,\xi$) follows
\begin{align}
y^{t}  &  =\Gamma(X+(I-M)\nu^{t})+By^{t-1}\label{Y}\\
\overline{y}^{t}  &  =\Gamma X+B\overline{y}^{t-1}, \label{Ybar}%
\end{align}
from which we obtain:
\[
\eta^{t}=\Gamma(I-M)\nu^{t}+B\eta^{t-1}%
\]
Since the $\nu_{i}^{t}$ are independent random variables, the evolution of the
vector of covariances follows:
\begin{equation}
w^{t}=\Lambda+\overline{B}w^{t-1} \label{w}%
\end{equation}

The evolution of $\overline{y}^{t}$ coincides with the case where there is no
noise. Lemma~B7 below extends Lemma~B2, showing that $\overline{H}\equiv
\sum_{k\geq0}\overline{B}^{k}$ (or the inverse $(I-\overline{B})^{-1}$) are
well-defined, which implies that $w^{t}$ has a well-defined limit
\begin{equation}
w=\overline{H}\Lambda, \label{limits}%
\end{equation}
\smallskip

\textbf{Lemma B7:} \textit{For }$K$\textit{ large enough, }$\overline{B}^{K}%
$\textit{ satisfies the condition of Lemma~B1.\smallskip}

\textbf{Proof of Lemma B7.} We extend the notion of sequences and paths to
pairs $ij\in N^{2}$ (rather than individuals). For any sequence of pairs
$\overline{q}=(i_{1}j_{1},...,i_{K}j_{K})$ (or equivalently, any pair of
sequences $\overline{q}=(q^{1},q^{2})=((i_{1},...,i_{K}),(j_{1},...,j_{K}))$)
and any matrix $A=(A_{ij})_{ij}$, and we let $\overline{\pi}^{A}(\overline
{q})=\pi^{A}(q^{1})\pi^{A}(q^{2})$. We define a path $\overline{q}$ as a
sequence such that $\overline{\pi}^{A}(\overline{q})>0$.\smallskip

We apply the argument of Lemma~B2 to paths $\overline{q}$ of pairs rather than
paths $q$ of individuals. Let $\overline{C}=\overline{B}^{K}$. Call
$\overline{Q}_{ij}^{K}$ the set of paths $\overline{q}=(q^{1},q^{2})$ of
length $K$ that start from $ij$ (to some $hk$), $\overline{Q}_{i}^{K,i_{0}}$
those for which $q^{1}$ goes through $i_{0}$. We have
\[
\sum_{hk}\overline{C}_{ij,hk}\equiv\overline{\pi}^{B}(\overline{Q}_{ij}%
^{K})\leq(1-\underline{\gamma}m_{i_{0}})\overline{\pi}^{A}(\overline{Q}%
_{i}^{K,i_{0}})+\overline{\pi}^{A}(\overline{Q}_{i}^{K}\backslash\overline
{Q}_{i}^{K,i_{0}})<1
\]
hence $\overline{C}$ satisfies the condition of Lemma~B1, $I-\overline{C}$ has
an inverse, and so does $I-\overline{B}$.$\blacksquare\smallskip$

\textbf{Proof of Proposition 7.}

(i) Let $\overline{C}=\overline{B}^{K}$ and $\overline{D}=I+\overline
{B}+...+\overline{B}^{K-1}$. Repeated substitutions in (\ref{w}) yield
\[
w^{t}=\Lambda^{0}+\overline{C}w^{t-K}%
\]
where $\Lambda^{0}=\overline{D}\Lambda$. By Lemma B7, Lemma B1 applies to
$\overline{C}$, so convergence of $w^{t}$ to $w$ is ensured.$\smallskip$

(ii) We bound the loss $V_{i}$ induced by the idiosyncratic errors. Recall%
\[
\eta_{i}^{t}=\gamma_{i}(1-m_{i})\nu_{i}^{t}+(1-\gamma_{i})\eta_{i}%
^{t-1}+\gamma_{i}(1-m_{i})A_{i}\eta^{t-1}%
\]
This implies that for any $p\in\Delta_{n}$, there exists $q$ $\in\Delta_{n}$
such that:
\begin{equation}
p.\eta^{t}=q.\eta^{t-1}+\sum_{i}\gamma_{i}(1-m_{i})p_{i}\nu_{i}^{t}\text{ and
}\sum_{i}q_{i}\geq1-\underline{m} \label{EqV}%
\end{equation}
Define $\underline{V}^{t}=\min_{p\in\Delta_{n}}var(p.\eta^{t})$. Note that
$V_{i}^{t}\geq\underline{V}^{t}$. Since $var(q.\eta^{t-1})\geq(1-\underline{m}%
)^{2}\underline{V}^{t-1}$, Equality (\ref{EqV}) implies $\underline{V}^{t}%
\geq(1-\underline{m})^{2}\underline{V}^{t-1}+\frac{1}{n}\underline{\gamma}%
^{2}(1-\underline{m})^{2}\varpi^{0}$, which yields the desired lower bound at
the limit.$\smallskip$

(iii) We now re-examine Result 2. We consider the effect of $m_{i}$ on the
vector of covariances $w$ where $w_{jk}=\lim E(y_{j}^{t}-\overline{y}_{j}%
^{t})(y_{k}^{t}-\overline{y}_{k}^{t})$. Recall $w=\Lambda+\overline{B}w$.
Since $\Lambda$ and $\overline{B}$ are non-increasing in $m_{i}$ and
$\Lambda_{ii}$ is strictly decreasing in $m_{i}$, $w_{ii}$ strictly decreases
with $m_{i}$, and $w$ is non-increasing in $m_{i}$. Combining all steps, over
the range $m_{i}<\underline{m}$, $L_{i}=\overline{L}_{i}+w_{ii}$ strictly
decreases with $m_{i}$, and $\sum_{k}L_{k}$ also strictly decreases with
$m_{i}$.$\blacksquare$\bigskip

\textbf{Proof of Result 6.} \textit{In addition to item (i) and (ii), we shall
prove the following statement: }(iii) \textit{If the lower bound
}$\underline{\gamma}$\textit{ on the choice set is sufficiently low and
}$\gamma_{i}=\underline{\gamma}$\textit{, }$V_{i}\leq1/\mid\log
\underline{\gamma}\mid$\textit{ for all }$m\geq\underline{m}$\textit{ and
}$\gamma$\textit{ within the choice set}.

Let $\overline{\gamma}=\max\gamma_{i}$ and recall:
\begin{equation}
w_{ij}=\sum_{h,k}B_{ih}B_{jk}w_{hk}+\Lambda_{ij} \label{wij}%
\end{equation}
where $\Lambda_{ij}=0$ if $i\neq j$ and $\Lambda_{ii}=(1-m_{i})^{2}(\gamma
_{i})^{2}\varpi^{0}$, and $B_{ii}=1-\gamma_{i}$, $B_{ij}=\gamma_{i}%
A_{ij}(1-m_{i})$.

The proof starts by proving item (i), that is, computing a uniform upper bound
on all $w_{ij}$ of the form (see step 1)
\begin{equation}
w_{ij}\leq c\overline{\gamma} \label{bound}%
\end{equation}
To prove (ii), we define $\widehat{w}=(w_{ij})_{j}$ as the vector of
covariances involving $i$, and show that there exists a matrix $C$ for which
$\sum_{k}C_{jk}\leq1$ for all $j$ and such that
\begin{equation}
\widehat{w}\leq(1-\underline{m})C\widehat{w}+\Gamma\label{C}%
\end{equation}
where $\Gamma_{j}\leq dp_{ij}$ for some $d$, with $p_{ij}=\gamma_{i}%
/(\gamma_{i}+\gamma_{j})$. This in turn implies that $\max_{j}w_{ij}\leq
\max_{j}\Gamma_{i}/\underline{m}$, which will prove (ii) (see step 3).

Finally, to prove (iii), we consider two cases. Either $\overline{\gamma}$ is
\textquotedblleft small" and (\ref{bound}) applies, or we can separate
individuals into a subgroup $J$ where all have a small $\gamma_{j}$, and the
rest of them with significantly larger $\gamma_{j}$. In the latter case, we
redefine $\widehat{w}=(w_{jk})_{j\in J,k}$ as the vector of covariances
involving some $j\in J$, and obtain inequality (\ref{C}) with $\Gamma_{jk}\leq
dp_{jk}$ for $k\notin J$ and $\Gamma_{jk}\leq d\gamma_{j}$ for $k\in J$, for
some $d$. By definition of $J$, all $\gamma_{j}$ and $p_{jk}$ are small, and
all $\Gamma_{jk}$ are thus small, which will prove (iii). Details are
below.\smallskip

\textbf{Step 1 }(item (i)) $w_{ij}\leq c\overline{\gamma}$ with $c=\varpi
^{0}/\underline{m}$. \smallskip

Let $\overline{V}=\max_{i}w_{ii}$ and $\overline{w}=\max_{i,j\neq i}w_{ij}$
and $\overline{w}=\max w_{i}$. For all $j\neq i$, $w_{ij}$ is a weighted
average between all $w_{h,k}$ and $0$, so $w_{ij}<\max(\overline{w}%
,\overline{V})$, hence $\overline{w}<\max(\overline{w},\overline{V})$, which
thus implies $\overline{w}\leq\overline{V}$. Consider $i$ that achieves
$\overline{V}$. Since $\sum_{h,k}B_{ih}B_{ik}=(1-\gamma_{i}m_{i})^{2}$, we
have:%
\begin{align*}
\overline{V}  &  =w_{ii}\leq(1-\gamma_{i}m_{i})^{2}\overline{V}+\gamma_{i}%
^{2}(1-m_{i})^{2}\varpi^{0}\text{ hence}\\
\overline{V}  &  \leq\frac{\gamma_{i}(1-m_{i})^{2}}{m_{i}}\varpi^{0}\leq
\frac{\varpi^{0}\overline{\gamma}}{\underline{m}}%
\end{align*}

\textbf{Step 2}. Let $p_{ij}=\gamma_{i}/(\gamma_{i}+\gamma_{j})$ and
$\overline{v}=2(c\overline{\gamma}+\omega_{0})$. We have:%
\begin{align}
w_{ii}  &  \leq\gamma_{i}p_{ii}\overline{v}+(1-\underline{m})\sum_{k}%
A_{ik}w_{ik}\label{step2a}\\
w_{ij}  &  \leq\gamma_{j}p_{ij}\overline{v}+(1-\underline{m})(p_{ij}\sum
_{k}A_{ik}w_{kj}+p_{ji}\sum_{k}A_{jk}w_{ik}) \label{step2b}%
\end{align}
These inequalities are obtained by solving for $w_{ij}$ in equation
(\ref{wij}), that is, we write
\[
(1-B_{ii}B_{jj})w_{ij}=\Gamma_{ij}+\sum_{k\neq i}B_{ii}B_{jk}w_{ik}%
+\sum_{k\neq i}B_{jj}B_{ik}w_{kj}+\sum_{k\neq i,h\neq j}B_{jk}B_{ih}w_{kj}.
\]
Observing that $2B_{ii}B_{ik}/(1-B_{ii}B_{jj})\leq(1-m_{i})A_{jk}$,
$B_{ii}B_{jk}/(1-B_{ii}B_{jj})\leq(1-m_{j})p_{ji}A_{jk}$, and $B_{jk}%
B_{ih}/(1-B_{ii}B_{jj})\leq2\gamma_{j}p_{ij}A_{jk}A_{ih}$ and $\Gamma
_{ii}/(1-B_{ii}B_{jj})\leq\gamma_{i}\omega^{0}$ yields (\ref{step2a}%
-\ref{step2b}).\smallskip

\textbf{Step 3} (item (ii))\textbf{.} It is immediate from (\ref{step2a}%
-\ref{step2b}) that (\ref{C}) holds with $C_{jk}\equiv A_{jk}$ and $\Gamma
_{j}=p_{ij}\gamma_{j}\overline{v}+p_{ij}c\overline{\gamma}\leq p_{ij}%
\overline{\gamma}(\overline{v}+c)\leq d\gamma_{i}$ for all $j$, for some $d$,
which permits to conclude that $\widehat{w}\leq d\gamma_{i}/\underline{m}%
$.\smallskip

\textbf{Step 4} (item (iii))\textbf{.} Let $\varepsilon=\frac{1}{K\mid
Log\underline{\gamma}\mid}$ with $K=5\varpi^{0}/\underline{m}^{2}$ and set
$\gamma_{i}=\underline{\gamma}$. Let us reorder individuals by increasing
order of $\gamma_{j}$. Consider first the case where $\gamma_{j+1}\leq
\gamma_{j}/\varepsilon$ for all $j=1,...,N-1$. Then $\overline{\gamma
}<\underline{\gamma}/\varepsilon^{N-1}$, and for $\underline{\gamma}$ small
enough, $\underline{\gamma}/\varepsilon^{N-1}<\varepsilon$, so $V_{i}\leq
c\varepsilon<1/\mid Log\underline{\gamma}\mid$.

Otherwise, there exists $j_{0}$ such that $\gamma_{j}\leq\underline{\gamma
}/\varepsilon^{j_{0}-1}$ for all $j\in J$, and $\gamma_{k}>\gamma
_{j}/\varepsilon$ for all $k\notin J$ and $j\in J$. It is immediate from
(\ref{step2a}-\ref{step2b}) that (\ref{C}) holds with $\Gamma$ such that, for
any $j\in J$,
\begin{align*}
\Gamma_{jk}  &  =\gamma_{j}\overline{v}\text{ if }k\in J\text{ and}\\
\Gamma_{jk}  &  =\gamma_{j}\overline{v}+p_{jk}\sum_{h\notin J}A_{jh}%
w_{hk}\text{ if }k\notin J
\end{align*}
By definition of $J$, for all $j\in J$, $\gamma_{j}\leq\underline{\gamma
}/\varepsilon^{N-1}<\varepsilon$ and for all $k\notin J$, $p_{jk}%
\leq\varepsilon$, which further that all $\Gamma_{jk}$ are bounded by
$\varepsilon(\overline{v}+c)\leq\underline{m}/\mid Log\underline{\gamma}\mid$,
which concludes the proof.$\blacksquare$\medskip

\subsection*{B.4 Proof of Proposition 13}

For fixed $x,\xi$, let $Y_{i}^{t}=(y_{i}^{t-k})_{k=0,..,K}$ denote the column
vector of $i$'s past recent opinions, and $Y^{t}=(Y_{i}^{t})_{i}$. One can
write $Y^{t}=X+BY^{t-1}$. $Y^{t}$ converges for standard reasons, to some
uniquely defined $Y$. Consider now the vector $y$ solution to
\[
y_{i}=m_{i}x_{i}+(1-m_{i})A_{i}(y+\xi_{i})
\]
and let $Y_{i}=(y_{i},...,y_{i})$ and $Y=(Y_{i})_{i}$. By construction, under
this profile of opinions, it does not matter when $i$ heard from $j$ because
opinions do not change. $Y$ thus solves $Y=X+BY$ and it coincides with $Y$.
The limit expected opinion vector under FJ is thus independent of the
communication protocol.$\blacksquare$\medskip

\subsection*{B.5 Coarse communication}

Recall $f$ is the fraction of agents choosing $a=0$, and call $y=\phi(f)$ the
associated "population opinion". We now consider two cases:

\textbf{Case 1}: $m=0$. Set $\xi>0$ and assume $f>0$. Each makes an inference
$z_{i}$ at least equal to $y+\xi$ regarding neighbors' opinions, so
eventually, under DG, each player of type $b_{i}$ may only report $0$ if
$b_{i}+y+\xi<0$. Under the large number approximation, a fraction at most
equal to $f^{\prime}=h(y+\xi)<f$ reports $0$, hence the fraction of agents
reporting $0$ eventually vanishes.

\textbf{Case 2}: $m$ small. When $m>0$, agents with signal $x_{i}$ believe the
state is $mx_{i}+(1-m)(y+\xi)$, which generates, under the large number
approximation, a fraction $f=Eh(mx_{i}+(1-m)(y+\xi))$ choosing $a=0$. The
long-run opinion $y$ thus solves%
\[
y=h^{-1}(Eh(m(\theta+\delta_{i})+(1-m)(y+\xi)))
\]
Call $\widehat{\xi}=y-\theta$ the resulting population estimation error. When
$m$ is small, $h$ is locally linear, so, since $E\delta_{i}=0$, $y\simeq
h^{-1}h(m(\theta+\frac{1-m}{m}\xi)+(1-m)y)$, which implies $\widehat{\xi
}\simeq\frac{1-m}{m}\xi$.

Assume now that player chooses $m_{i}$ while others choose $m$. For player
$i$, the estimation error is $\Delta_{i}\equiv m_{i}\delta_{i}+(1-m_{i}%
)(\widehat{\xi}+\xi)\simeq m_{i}\delta_{i}+(1-m_{i})\frac{\xi}{m}$. Assuming
that $\theta$ is drawn from a flat distribution with large support, the
expected loss $L_{i}(\Delta)$ from estimating $\theta$ with an error
$\Delta_{i}$ is quadratic in $\Delta_{i}$ and independent of $b_{i}%
$,\footnote{When $\Delta_{i}>0$, $L(\Delta_{i})=\int_{-\Delta_{i}-b_{i}%
}^{-b_{i}}-(\theta+b_{i})d\theta=\frac{\Delta^{2}}{2}.$} so $L(\Delta)$ is
proportional to the variance of the error that $i$ makes. To minimize the
variance of $\Delta_{i}$, player $i$ sets $m_{i}=\frac{\varpi}{m^{2}}$, so in
equilibrium $m^{\ast}=\varpi^{1/3}$.

Regarding the social optimum, when all choose $m$, the estimation error is
$m\delta_{i}+(1-m)\frac{\xi}{m}$. For $\varpi$ small, the variance of this
error is minimized for $m\simeq(2\varpi^{1/4})$.\medskip

\subsection*{B.6 An alternative modeling of errors}

To conclude this Appendix, we briefly comment on our modeling of errors. Given
the way we index errors, it is natural to interpret $\xi_{i}$ as a persistent
error that $i$ makes in \textit{processing} or hearing others' opinions. We
discuss below an alternative model where $i$ does not make processing errors
but makes a persistent error $\zeta_{i}^{e}$ in \textit{expressing her
opinion}. In this case Equation~(\ref{eqz}) becomes%
\[
z_{i}^{t}=A_{i}(y^{t}+\zeta^{e})
\]
so in effect, $i$ is subject to an error $\xi_{i}\equiv A_{i}\zeta^{e}$. Our
analysis thus extends to this alternative modelling with $\xi_{i}$
appropriately re-defined. With perfectly correlated errors, this alternative
modeling yields $\xi_{i}=\zeta_{i}^{e}$, so the analysis is unchanged. With
independent errors, the errors $\xi_{i}$ (hence the cumulated errors
$\widehat{\xi}_{i}$) now potentially depends on the network structure. We
re-examine our three network examples in light of this alternative modeling.
Specifically, we compare the cumulated errors terms when agents are subject to
processing errors ($\xi_{i}=\xi_{i}^{p}$) (respectively expressing errors
$\xi_{i}=A_{i}\zeta^{e}$), and denote by $\widehat{\varpi}_{i}^{p}$ and
$\widehat{\varpi}_{i}^{e}$ the respective variances, assuming that all errors
$\xi_{i}^{p}$ and $\zeta_{i}^{e}$ are independent and homogenous. We let
$\varpi=var\xi_{i}^{p}=var\zeta_{i}^{e}$. We have\smallskip

\textbf{Proposition~B1:} \textit{For the directed circle, }$\widehat{\varpi
}_{i}^{p}=\widehat{\varpi}_{i}^{e}$\textit{. For the complete network,
}$\widehat{\varpi}_{i}^{p}=\widehat{\varpi}_{i}^{e}+\frac{n-2}{n-1}\varpi
$\textit{. For the star network, }$\widehat{\varpi}_{0}^{p}=\widehat{\varpi
}_{0}^{e}+\frac{(2-m)(n-1)}{mn}\varpi$\textit{.}\smallskip

The main insight of this Proposition is that although the magnitude of the
one-shot error $\xi_{i}$ that a player faces may differ substantially
depending on whether we consider processing or expressing errors, the
cumulated error terms do not differ much in the sense that terms of order
$\varpi/m^{2}$ remain the same.\footnote{For example, in a star network,
$\xi_{0}=\overline{\zeta}^{e}$ so $var\xi_{0}=\varpi/n$ for expressing errors,
and $var\xi_{0}=\varpi_{0}=\varpi$ for processing errors. Nevertheless, the
cumulated errors are respectively $(\overline{\zeta}^{e}+(1-m)\zeta_{0}%
^{e})/m$ and $(\xi_{0}+(1-m)\overline{\xi})/m$.} The consequence is that,
while processing errors generate slightly larger cumulated errors than
communication errors, the effect is negligible for small $\varpi$, and at
least for the specific networks considered above, equilibrium analysis is then unchanged.

\textbf{Proof of Proposition B1:} Lemma B3 to B6 provide cumulated error terms
for processing errors. We use these Lemma to derive the cumulated error terms
for expressing errors, using $\xi_{i}\equiv A_{i}\zeta^{e}$. For the directed
circle, $\xi_{i}=\zeta_{i+1}^{e}$, so we immediately obtain $\widehat{\varpi
}_{i}^{p}=\widehat{\varpi}_{i}^{e}$. For the full network, $\xi_{i}%
=\overline{\zeta}_{-i}^{e}$, so $\overline{\xi}_{-i}=\frac{1}{n-1}\zeta
_{i}^{e}+\overline{\zeta}_{-i}^{e}(1-\frac{1}{n-1})$, which further implies
$\widehat{\xi}_{i}^{e}=\frac{1-m}{m(n-1)}\zeta_{i}^{e}+\frac{1}{m}%
\overline{\zeta}_{-i}^{e}$, hence the desired comparison. For the star
network, $\xi_{i}=\zeta_{0}^{e}$ and $\xi_{0}=\overline{\zeta}^{e}$, so
$\widehat{\xi}_{0}^{e}=\frac{1-m}{m}\zeta_{0}^{e}+\frac{1}{m}\overline{\zeta
}^{e},$ hence the desired comparison. Note that, for the cumulated errors
faced by peripherical players, one can compute $\widehat{\varpi}_{i}^{p}$ and
$\widehat{\varpi}_{i}^{e}$ for fixed $\rho_{0}$. In equilibrium, for small
$\varpi$, omitting terms of higher orders, one can check that in equilibrium,
$\widehat{\varpi}^{p}-\widehat{\varpi}^{e}\simeq(1-\frac{1}{n^{2}}%
)\varpi/m^{\ast}$ with $m^{\ast}\simeq((1+1/n)\varpi)^{1/3}$.$\blacksquare$
\end{document}